\documentclass[graybox, nosecnum]{svmult}

\usepackage{mathptmx}       % selects Times Roman as basic font
\usepackage{helvet}         % selects Helvetica as sans-serif font
\usepackage{courier}        % selects Courier as typewriter font
\usepackage{type1cm}        % activate if the above 3 fonts are
\usepackage{makeidx}         % allows index generation
\usepackage{graphicx}        % standard LaTeX graphics tool
\usepackage{multicol}        % used for the two-column index
\usepackage[bottom]{footmisc}% places footnotes at page bottom
\usepackage{hyperref}        %for hyperlinks
\usepackage{soul}            % for high-lighting of text

\usepackage{amsmath}
\usepackage{amssymb}
\usepackage{caption}
\usepackage{subcaption}

\newcommand{\bra}[1]{\langle#1|}
\newcommand{\ket}[1]{|#1\rangle}

\newcommand{\commut}[1]{\left[#1\right]}

\makeindex             % used for the subject index
\begin{document}
%\tableofcontents{}
\title*{\textit{Ab Initio} Nuclear Reaction Theory with Applications to Astrophysics}
\author{Petr Navr\'atil\thanks{corresponding author} and Sofia Quaglioni}
\institute{Petr Navr\'atil \at TRIUMF, 4004 Wesbrook Mall, Vancouver, British Columbia V6T 2A3, Canada, \email{navratil@triumf.ca}
\and Sofia Quaglioni \at Lawrence Livermore National Laboratory, P.O. Box 808, L-414, Livermore, California 94551, USA, \email{quaglioni1@llnl.gov}}
\maketitle
\abstract{We present an introduction to \textit{ab initio} nuclear theory with a focus on nuclear reactions. After a high-level overview of \textit{ab initio} approaches in nuclear physics, we give a more detailed description of the no-core shell model technique equivalent to a large extent to configuration-interaction methods applied in quantum chemistry. We then introduce the no-core shell model with continuum approach that provides a quantum many-body description of nuclear reactions. After a brief review of nuclear reactions important for astrophysics, we present examples of results of \textit{ab initio} calculations of radiative capture and transfer reactions in light nuclei.}

\section{\textit{Ab Initio} Nuclear Theory}\label{abinitio}
\subsection{Introduction}\label{Intro}

Understanding the structure and the dynamics of atomic nuclei as many-body systems of protons and neutrons interacting through the strong, electromagnetic, and weak forces is one of the central goals of nuclear physics. One of the major reasons why this goal has yet to be accomplished lies in the complex nature of the strong nuclear force, emerging form the underlying theory of Quantum Chromodynamics (QCD). At the low energies relevant to the structure and dynamics of nuclei, QCD is non-perturbative and very difficult to solve. The relevant degrees of freedom for nuclei are nucleons, i.e., protons and neutrons, that are not fundamental particles but rather complex objects made of quarks, antiquarks and gluons. Consequently, the strong interactions among nucleons is only an ``effective" interaction emerging non-perturbatively from QCD. At present, our knowledge of the nucleon-nucleon (NN) interactions is limited to models. The most advanced and most fundamental of these models rely on a low-energy effective field theory (EFT) of  
QCD, chiral EFT~\cite{Weinberg1991}. This theory is built on the symmetries of QCD, most notably the approximate chiral symmetry. Chiral EFT involves unknown parameters, low-energy constants (LECs) that in principle can be calculated within QCD, e.g., using the lattice QCD approach, but currently they are fitted to experimental data. Chiral EFT predicts higher-body forces, in particular a three-nucleon (3N) interaction that plays an important role in nuclear structure and dynamics. 

{\it Ab initio} calculations in nuclear physics use nucleons as the relevant degrees of freedom, start from the realistic forces among nucleons, recently almost exclusively the chiral EFT interactions that describe accurately the two-nucleon system and three-nucleon bound states, and aim at predicting the properties of atomic nuclei. Solving the {\it ab initio} many-body problem is a very challenging task because of the complex nature of nuclear forces and because of our limited knowledge of these forces. The high-level strategy is to solve the non-relativistic many-nucleon Schr\"{o}dinger equation with the inter-nucleon interactions as the only input. This can be done exactly for the lightest nuclei ($A{=}3,4$) \cite{FRIAR19934,Nogga199719,Barnea2001565,PhysRevC.64.044001}. However, using new methods, including those that scale polynomially with the basis size, combined with high-performance computing and well-controlled approximations, {\it ab initio} calculations have recently progressed tremendously and become applicable to nuclei as heavy as $^{100}$Sn~\cite{Gysbers2019NatPhys} and beyond. 

The progress of {\it ab initio} theory has been in particular quite dramatic concerning the description of bound-state properties of nuclei. For light nuclei, the Green's function Monte Carlo method (GFMC) \cite{Pudliner1995,annurev.nucl.51.101701.132506,PhysRevC.66.044310} has been applied up to $A\leq 12$. The no-core shell model (NCSM) \cite{PhysRevLett.84.5728,BARRETT2013131} with its importance-truncated extension \cite{Roth2007,Roth2009} allows the {\it ab initio} description of the static properties of nuclei up to oxygen isotopes \cite{Hergert2013}. Other NCSM extensions, e.g., the symmetry-adapted NCSM \cite{PhysRevLett.111.252501} and no-core Monte-Carlo shell model \cite{PhysRevC.86.054301} are under active development. The symmetry-adapted (SA) NCSM is capable to describe collective excitations in p- and sd-shell nuclei~\cite{PhysRevLett.124.042501}. The NCSM was applied in studies of order-by-order convergence of chiral EFT interactions~\cite{PhysRevC.103.054001}. Recently, methods such as the coupled cluster (CCM) \cite{PhysRevLett.94.212501,PhysRevLett.104.182501,Binder2013,PhysRevLett.109.032502,0034-4885-77-9-096302,PhysRevC.91.064320}, the self-consistent Green's function (SCGF) \cite{Cipollone2013} and its Gorkov generalization \cite{PhysRevC.89.061301}, and the newly developed in-medium similarity renormalization group (IM-SRG) method \cite{PhysRevLett.106.222502,Hergert2013,Hergert2013a,PhysRevC.90.041302,PhysRevLett.113.142501} achieved high accuracy and predictive power for nuclei up to the tin region with a full capability to use chiral NN+3N interactions. The valence-space version of the IM-SRG method~\cite{annurev-nucl-101917-021120} can reach all nuclei accessible by a standard shell model while enabling the type of predictive power characteristic of {\it ab initio} methods, as demonstrated by their predictions of drip lines from helium to iron~\cite{PhysRevLett.126.022501}. The IM-SRG was also combined with the NCSM to extend its reach of applicability~\cite{PhysRevLett.118.152503}. Another very recent and important development is the introduction of emulators of many-body calculations that can facilitate the uncertainty quantification of {\it ab initio} methods~\cite{PhysRevLett.123.252501}. Further, there has been progress in Monte Carlo methods such as the nuclear lattice EFT \cite{PhysRevLett.109.252501,PhysRevLett.112.102501} as well as the auxiliary-field diffusion Monte Carlo (AFDMC) method and the GFMC that are now also able to use chiral EFT NN+3N interactions that had been previously out of reach because of their non-locality~\cite{PhysRevC.90.054323,Lynn:2015jua}.

As for the description of the dynamics in the continuum of energy, for $A{=}3,4$ systems there are several successful exact methods, e.g., the Faddeev~\cite{PhysRevC.63.024007}, Faddeev-Yakubovsky ~\cite{PhysRevC.70.044002,PhysRevC.79.054007}, Alt-Grassberger and Sandhas (AGS)~\cite{PhysRevC.75.014005,PhysRevLett.98.162502}, and hyperspherical harmonics (HH)~\cite{0954-3899-35-6-063101,Marcucci2009} methods. For $A>4$ nuclei, concerning calculations of nuclear resonance properties, scattering and reactions, there has been less activity and the no-core shell model with resonating-group method (NCSM/RGM) \cite{Quaglioni2008,Quaglioni2009} and in particular the no-core shell model with continuum (NCSMC) method \cite{Baroni2013,Baroni2013L} discussed later in this chapter are cutting edge approaches. Still the field is rapidly evolving also in this area. Monte Carlo calculations~\cite{Nollett2007,Lynn:2015jua} and the Faddeev-Yakubovsky method~\cite{PhysRevC.97.044002} were applied to calculate $n{-}^4$He scattering, nuclear lattice EFT calculations were applied to the $^4$He-$^4$He scattering \cite{Elhatisari:2015iga}, and the $p{-}^{40}$Ca scattering was calculated within the CCM with the Gamow basis \cite{Hagen2012}. The CCM with the Gamow basis was also used to investigate resonances in $^{17}$F~\cite{PhysRevLett.104.182501} and in oxygen isotopes~\cite{Hagen2012}. The symmetry-adapted NCSM approach has been applied to study alpha clustering and the capability to describe scattering is under development~\cite{annurev-nucl-102419-033316}.

\subsection{Hamiltonian}\label{Hamiltonian}

{\em Ab initio} approaches start from the microscopic Hamiltonian for the $A$-nucleon system 
\begin{equation}\label{eq:ham_Ham}
\hat{H} = \hat T_{\rm int} +\hat V
\end{equation}
composed of the intrinsic kinetic energy operator $\hat T_{\rm int}$ and the nuclear interaction $\hat V = \hat V^{NN}+\hat V^{3N}+\dots$, which describes the strong, electro-magnetic, and--in special applications--also the weak interaction among nucleons. The interaction $\hat{V}$ typically consists of realistic NN and 3N contributions that accurately reproduce few-nucleon properties, but in principle can also contain higher many-nucleon contributions. More specifically, the Hamiltonian can be written as
\begin{equation}
\hat{H}=\frac{1}{A}\sum_{i<j=1}^A\frac{(\hat{\vec{p}}_i-\hat{\vec{p}}_j)^2}{2m}+\sum_{i<j=1}^A \hat{V}^{NN}_{ij}+\sum_{i<j<k=1}^A \hat{V}^{3N}_{ijk}+\dots,
\label{H}
\end{equation} 
where $m$ is the nucleon mass and $\vec{p}_i$ the momentum of the $i$th nucleon.
The electro-magnetic interaction is typically described by the Coulomb force, while the determination of the strong interaction poses a tremendous challenge.  

According to the Standard Model, the strong interaction between nucleons is described by QCD with quarks and gluons as fundamental degrees of freedom. 
However, the nuclear structure and dynamics phenomena we are investigating are dominated by low energies where QCD becomes non-perturbative. This so far impedes a direct derivation of the nuclear interaction from the underlying theory. 

\begin{figure}
\centering\includegraphics[width=1.0\textwidth]{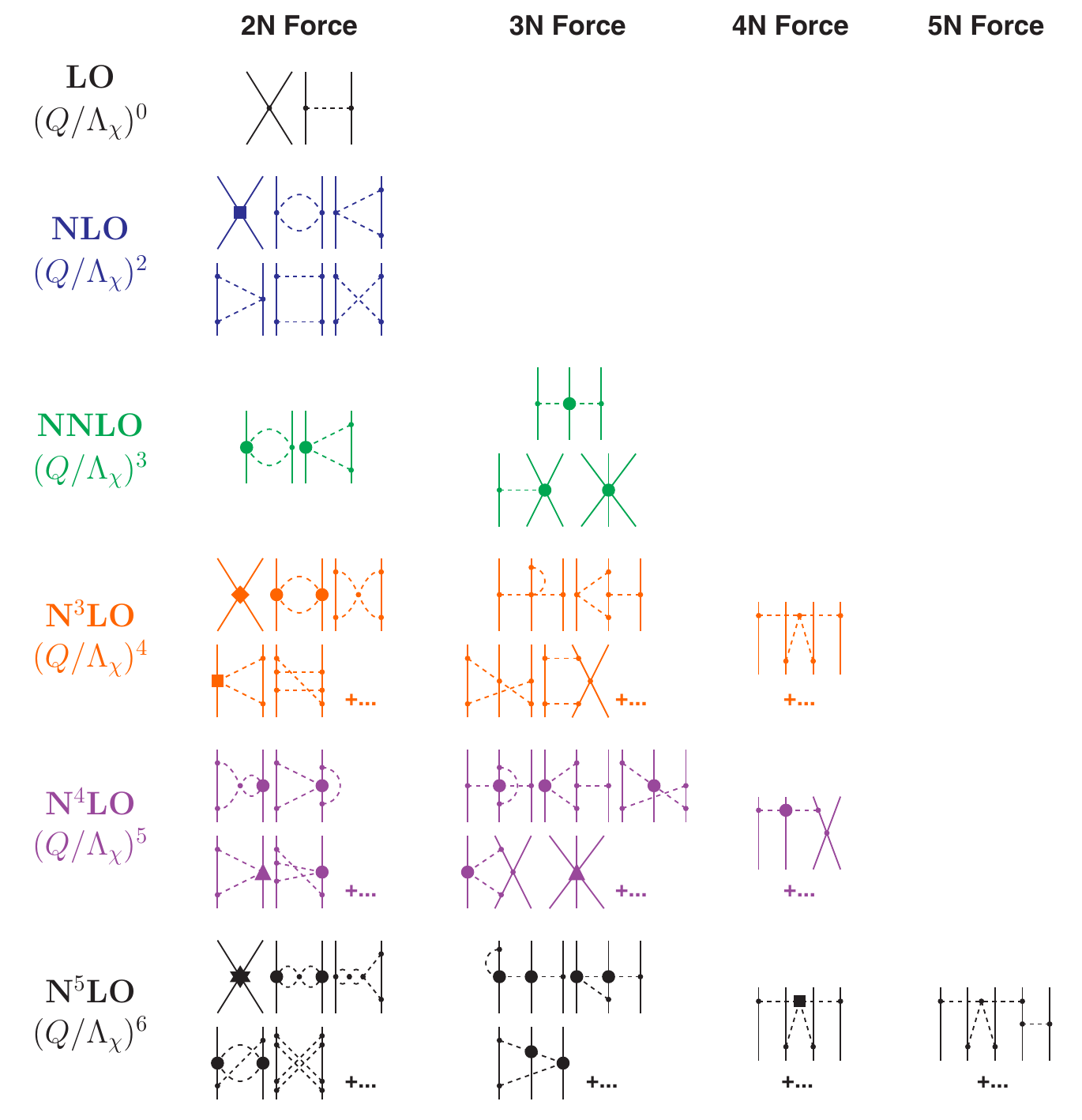}
\caption{{Hierarchy of nuclear forces in chiral EFT~\cite{Entem2017}:} 
{\small The interaction diagrams up to next-to-next-to-next-to-next-to-next-to leading order (N$^{5}$LO) arranged by the particle rank of the interaction. Dashed lines represent pions and the solid lines nucleons. Small dots, large solid dots, solid squares, triangles, diamonds, and stars denote vertices at increasing expansion orders. Figure from Ref.~\cite{Entem2017} where further details are given.
}
\label{fig:XEFTDiagram}}
\end{figure}

Inspired by basic symmetries of the Hamiltonian and the meson-exchange theory proposed by Yukawa~\cite{Yukawa35}, phenomenological high-precision NN interactions, such as the Argonne V18~\cite{Wiringa1995} and CD-Bonn~\cite{Mach01} potentials, have been developed. These interactions provide an accurate description of NN systems, but sizeable discrepancies are observed in nuclear structure applications to heavier nuclei ~\cite{CaNa02,NaOr02,PhysRevC.66.044310}. This indicates the importance of many-nucleon interactions beyond the two-body level and reveal the necessity for a consistent scheme to construct the nuclear interactions.
Thus, Weinberg formulated an effective theory for the low-energy regime using nucleons and
pions as explicit degrees of freedom~\cite{Weinberg79}. Chiral EFT~\cite{Weinberg1990,Weinberg1991} uses a low-energy expansion illustrated in Fig.~\ref{fig:XEFTDiagram} in terms of $(Q/\Lambda_{\chi})^{\nu}$, where $Q$ relates to the nucleon momentum/pion mass and $\Lambda_{\chi}$ corresponds to the break down scale of the chiral expansion, that is typically on the order of $1\,\text{GeV}$. This expansion allows for a systematic improvement of the potential by an increase of the chiral order $\nu$.
Moreover, the chiral expansion  provides a hierarchy of NN, 3N, and many-nucleon interactions in a consistent scheme~\cite{OrRa94,VanKolck94,EpNo02,Epelbaum06}. Chiral EFT consistency implies that the same LECs may appear in NN, 3N, as well as other expansion terms including the electro-magnetic and weak currents.
Since the chiral expansion is only valid at low energies it is necessary to suppress high-momentum contributions beyond a certain cutoff $\Lambda_{\text{cut}}$ by introducing a regularization. There are different possible choices for the regulator function and the cutoff, which determine the regularization scheme. Typically, the cutoff $\Lambda_{\text{cut}}$ varies in the range of 400-600 MeV. Most of the calculations presented later in this chapter use NN interactions from Refs.~\cite{Entem2003} and \cite{Entem2017} with $\Lambda_{\text{cut}}{=}500$~MeV. 

\subsection{Similarity Renormalization Group Transformations}\label{sec:srg}

Chiral interactions are already rather soft compared to phenomenological high-precision interactions such as the Argonne V18~\cite{Wiringa1995} and CD-Bonn~\cite{Mach01} potentials, owing to the regularization that suppresses high-momentum contributions, as described in Section Hamiltonian.
Nevertheless, most many-body methods--including those based on the NCSM we are focusing on here--cannot achieve convergence in feasible model spaces due to 
remaining short-range and tensor correlations induced by the chiral interactions. 
Therefore, additional transformations such as the unitary correlation operator method (UCOM)~\cite{Roth2010}, %the 
$V_{{\rm low-}k}$ renormalization group method~\cite{BoKu03,BoKu03b,BoFu07b} or the Okubo-Lee-Suzuki similarity transformation~\cite{SuLe80,Okub54} are used to soften the interactions. The most successful unitary transformation approach in nuclear 
physics is the similarity renormalization group (SRG)~\cite{Wegner1994,Bogner2007,SzPe00} that is presented in the following. This transformation provides a model-space and nucleus independent softened interaction and allows for consistent unitary transformations of the NN and 3N components of the interaction.   

The basic concept of the SRG is the first-order differential operator equation
\begin{equation} \label{eq:srg_floweq}
  \frac{d}{ds} \hat{H}_{s} = \big{[} {\hat{\eta}_{s}},{\hat{H}_{s}}\big{]}\,,
\end{equation}
that defines the continuous unitary transformation $\hat{H}_{s}=\hat{U}^{\,\dagger}_{s} \, \hat{H} \, \hat{U}_{s}$ with $\hat{H}$ given by Eq.~(\ref{eq:ham_Ham}), where the unitary operator $\hat{U}_{s}$ depends on the continuous flow parameter $s$.
In this flow equation $\hat{H}_{s}$ denotes the SRG evolved Hamiltonian depending on the flow parameter $s$ and the anti-Hermitian dynamic generator 
\begin{equation} \label{eq:srg_gen}
\hat{\eta}_{s} = -\hat{U}^{\,\dagger}_{s} \frac{d}{ds} \hat{U}_{s} = - \hat{\eta}^{\dagger}_{s}\,.   
\end{equation}
The typical choice for the generator used in the majority of nuclear structure and reaction applications is the commutator of the kinetic energy with the Hamiltonian, i.e.,
\begin{equation} \label{eq:srg_canonicgen}
  \hat{\eta}_{s} = \left(\frac{m}{\hslash^2}\right)^2\; \commut{\hat T_{\rm int},\hat{H}_{s}}\,,
\end{equation}
where $\hat T_{\rm int}$ constitutes the trivial fix point of the flow of the Hamiltonian, such that the high- and low-momentum contributions of the interaction decouple.
For this generator choice it is reasonable to associate the flow parameter with a momentum scale, using the relation $\Lambda=s^{-(1/4)}$ as often done in the literature~\cite{Jurgenson2009,Bogner201094}.

When aiming at observables other than binding and excitation energies it is formally necessary to transform the corresponding operators $ \hat{O}_{s}=\hat{U}^{\,\dagger}_{s} \, \hat{O} \, \hat{U}_{s}$, which can be achieved by evaluating $\hat{U}_{s}$ directly or by solving the flow equation
\begin{equation}\label{eq:srg_floweqO}
\frac{d}{ds} \hat{O}_{s} = \commut{\hat{\eta}_{s},\hat{O}_{s}}\, .
\end{equation}
Because the dynamic generator contains the evolved Hamiltonian, the flow equations for the operator $\hat{O}_{s}$ and the Hamiltonian $\hat{H}_s$ need to be evolved simultaneously.
We refer to Ref.~\cite{ScQu14,Schuster2015,Gysbers2019NatPhys} for recent applications and stress that there is work in progress to perform SRG transformations of observables. 

It is important to note that Eq.~\eqref{eq:srg_floweq} is an operator relation in  the $A$-body Hilbert space. Due to the repeated multiplication of the operators on the right hand side of the flow equation, irreducible many-body contributions beyond the initial particle rank of the Hamiltonian are induced. Generally, contributions beyond the three-body level cannot be considered for technical reasons.
This limitation causes one of the most challenging problems in context of the SRG transformation, since the unitarity is formally violated. Thus, it is necessary to confirm the invariance of physical observables under the transformation. In practice a variation of the flow parameter $s$ or $\Lambda$ is used as a diagnostic tool to access the impact of omitted many-body contributions.

For practical applications the flow equation \eqref{eq:srg_floweq} is represented in a basis and the resulting first-order coupled differential equations are solved numerically. Due to the simplicity of the evolution the SRG can be implemented in the three-body space and possibly even beyond. 
The most efficient formulation for the SRG evolution with regard to  descriptions of finite nuclei is performed in the Jacobi HO representation~\cite{Jurgenson2009,Jurgenson2011,Roth2011} using a subsequent Talmi-Moshinsky transformation~\cite{PhysRevC.5.1534,KaKa01} to the single-particle representation that is utilized by the many-body approaches, see Ref.~\cite{Roth2014} for a detailed explanation. 
There are also implementations of the three-body SRG evolution performed in other basis representations, such as the partial-wave decomposed momentum-Jacobi basis~\cite{Hebe12} and the hyperspherical momentum basis~\cite{Wend13}. However, so far only the SRG in the HO basis has been used to provide reliably evolved 3N interactions and operators for nuclear structure calculations beyond the lightest nuclei.      

It has been shown that the two-pion exchange part of the 3N interaction with a local regulator induces irreducible contributions beyond the three-body level that become sizeable in the mid-p shell~\cite{Roth2014}. As a consequence alternative formulations of the dynamic generator have been explored to avoid induced many-body contributions from the outset~\cite{Dicaire14}. In addition, it has been observed that a reduction of the 3N cutoff from 500~MeV to 400~MeV for 3N with a local regulator strongly suppresses the impact of induced many-body contributions~\cite{Roth2014} and allows for reliable applications beyond p- and sd-shell nuclei. For chiral 3N interactions with non-local regulators, the SRG induced four- and higher-body contributions appear much less significant~\cite{Huther2020,Soma2020,PhysRevC.103.035801}.

\section{No-Core Shell Model}
\label{NCSM}

Expansions on square integrable many-body states are among the most common techniques for the description of the static properties of nuclei. 
The {\it ab initio} NCSM~\cite{PhysRevLett.84.5728,PhysRevC.62.054311,BARRETT2013131} is one of such techniques.
Nuclei are considered as systems of $A$ non-relativistic point-like nucleons interacting through realistic inter-nucleon interactions discussed in Section Hamiltonian. All nucleons are active degrees of freedom. Translational invariance as well as angular momentum and parity of the system under consideration are conserved. 

\begin{figure}
    \centering
        \begin{subfigure}[b]{0.32\textwidth}
         \centering
         \includegraphics[width=\textwidth]{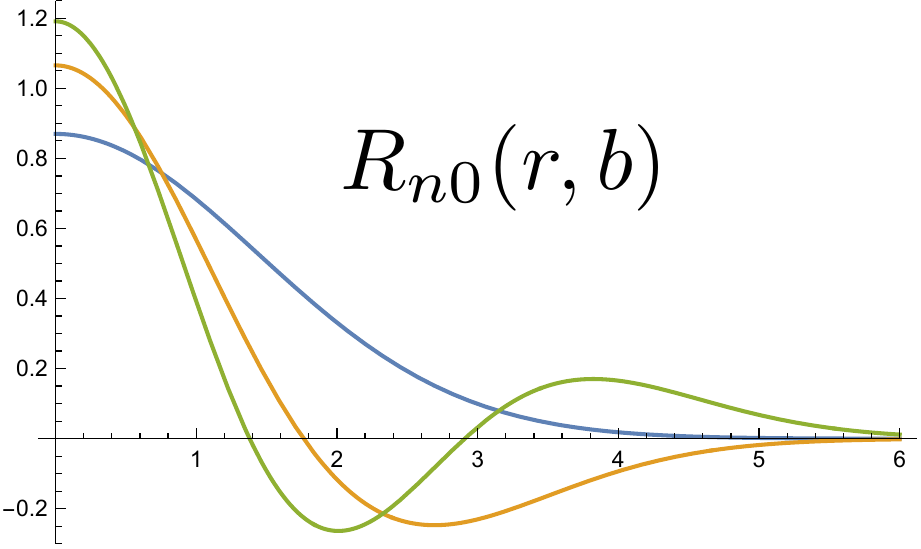}
%         \caption{$y=x$}
%         \label{fig:y equals x}
     \end{subfigure}
            \hfill
        \begin{subfigure}[b]{0.32\textwidth}
         \centering
         \includegraphics[width=\textwidth]{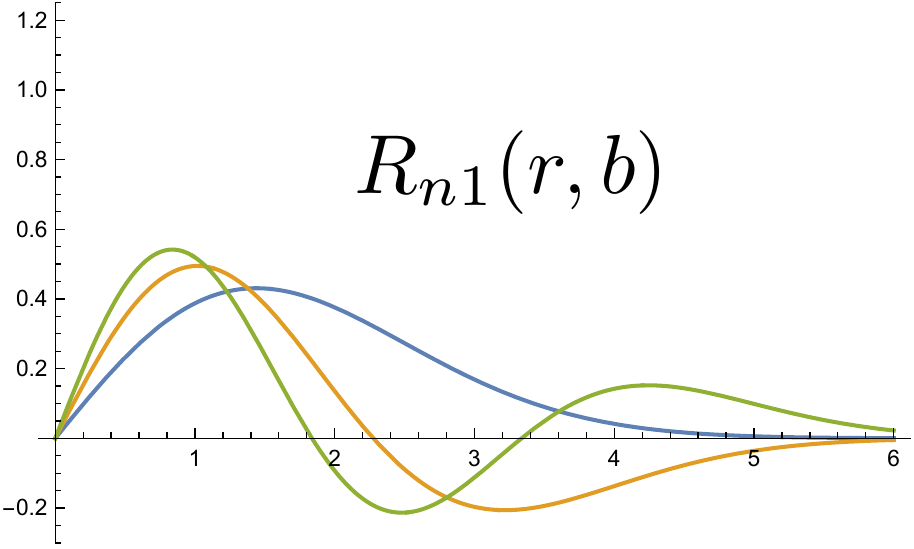}
%         \caption{$y=x$}
%         \label{fig:y equals x}
     \end{subfigure}
            \hfill
        \begin{subfigure}[b]{0.32\textwidth}
         \centering
         \includegraphics[width=\textwidth]{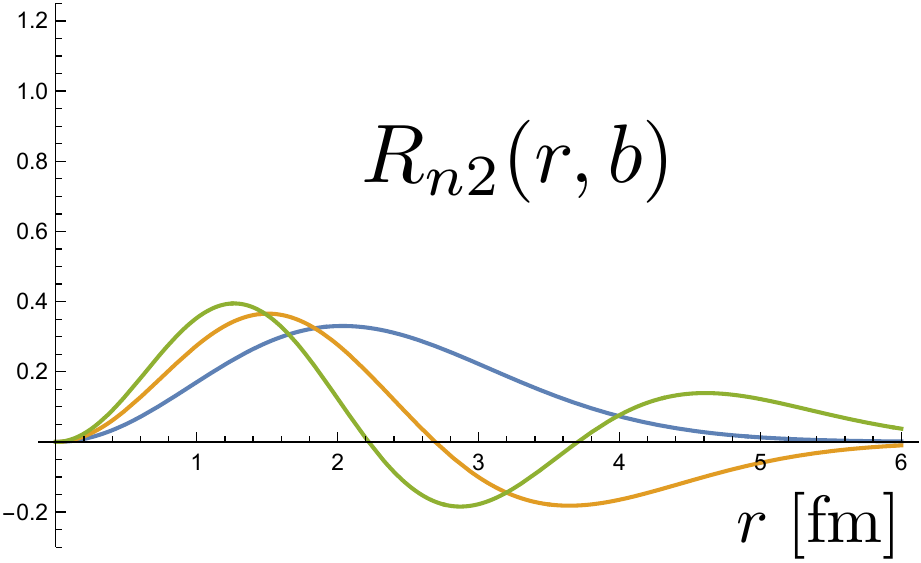}
%         \caption{$y=x$}
%         \label{fig:y equals x}
     \end{subfigure}
    \caption{Radial HO wave functions $R_{nl}(r,b)$, in fm$^{-3/2}$, for the radial quantum number $n{=}0,1,2$ that equals the number of nodes, the orbital momentum $l{=}0,1,2$, and $b{=}1.44$ fm.}
    \label{fig:Rnl}
\end{figure}

The many-body wave function is cast into an expansion over a complete set of antisymmetric $A$-nucleon HO basis states containing up to $N_{\rm max}$ HO excitations above the lowest possible configuration. The HO wave function depending on a single coordinate is expressed as $\varphi_{nlm}(\vec{r}){=}R_{nl}(r,b)Y_{lm}(\hat{r})$ with $R_{nl}$ the radial HO wave function and $Y_{lm}$ the spherical harmonics. The oscillator length $b$ (in fm) relates to the HO frequency by $b{=}b_0{=}\sqrt{\hslash / (m \Omega)}$ with $m$ the nucleon mass. We note that for a typical HO frequency used in NCSM calculations, $\hslash\Omega{=}20$ MeV, the oscillator length is equal to $b_0{=}1.44$ fm. The $n$ is the radial HO quantum number equal to the number of nodes of $R_{nl}(r,b_0)$, $l$ is the orbital momentum quantum number, and $m$ its projection on the quantization axis. With the positive value at origin convention, the $R_{nl}$ is given by
\begin{equation}
R_{nl}(r,b) = \sqrt{\frac{2\Gamma(n+1)}{(b^2)^{l+\frac{3}{2}}\Gamma(n+l+\frac{3}{2})}}\,r^{l} \exp\Big(-\frac{r^2}{2b^2}\Big)\,L_{n}^{l+\frac{1}{2}}\bigg(\frac{r^2}{b^2}\bigg) \,,    
\end{equation}
where $\Gamma$ is the gamma function and $L_{n}^{l+\frac{1}{2}}$ the generalized Laguerre polynomial. The HO radial wave functions for $n{=}0,1,2$, $l{=}0,1,2$, and $b{=}1.44$ fm are shown in Fig.~\ref{fig:Rnl}.

The HO basis may depend on either Jacobi relative~\cite{Jacobi_NCSM} or single-particle
coordinates~\cite{PhysRevC.62.054311}. In the former case, the wave function does not contain the center of mass (c.m.) motion, but antisymmetrization is complicated. In the latter case,  
antisymmetrization is trivially achieved using Slater determinants, but the c.m. degrees of freedom are included in the basis. In the simplest case of a two-nucleon system, the Jacobi coordinates are introduced by
\begin{subequations}\label{jacobiam11}
\begin{eqnarray}
\vec{\xi}_0 &=& \textstyle{\sqrt{\frac{1}{2}}}\left[\vec{r}_1+\vec{r}_2\right]
\; , \\
\vec{\xi}_1 &=& \textstyle{\sqrt{\frac{1}{2}}}\left[\vec{r}_1-\vec{r}_2\right] \; .
\end{eqnarray}
\end{subequations}
Here, $\vec{\xi}_0$ is proportional to the center of mass of the two-nucleon system: $\vec{R}^{(2)}_{\rm c.m.}=\sqrt{\frac{1}{2}}\vec{\xi}_0$. We are interested in the intrinsic motion of the nucleus. The basis functions of the two-nucleon system will then be given by
\begin{eqnarray}\label{A2relwf}
\langle \vec{\xi}_1 \sigma_1 \sigma_2 \tau_1 \tau_2|n l s j m t m_t\rangle &=& \sum_{m_l m_s} (l m_l s m_s | j m)\varphi_{nlm_l}(\vec{\xi}_1) (\chi(\sigma_1) \chi(\sigma_2))^{(s)}_{m_s} \nonumber \\ 
&\times&(\chi(\tau_1) \chi(\tau_2))^{(t)}_{m_t}    
\end{eqnarray}
with $\sigma_i$, $\tau_i$ the spin and isospin coordinates, respectively, $\chi$ the spinor functions, $s$ and $t$ is the total spin and isospin of the two nucleons, respectively, and $m_t{=}1,0,-1$ for the proton-proton, proton-neutron, neutron-neutron system, respectively. The antisymmetry of the wave function (\ref{A2relwf}) with respect of the exchange of the two nucleons is achieved by enforcing the condition $(-1)^{l+s+t}{=}-1$. However, a generalization to systems of more nucleons is non-trivial. A possible way to proceed is to diagonalize the antisymmetrizer, see, e.g., Ref.~\cite{Jacobi_NCSM} for details.

When using the single-particle coordinates, we introduce single-nucleon HO wave functions
\begin{eqnarray}\label{spHOwf}
   \langle \vec{r}_1 \sigma_1 \tau_1 |n_1 l_1 j_1 m_1 m_{t1}\rangle &\equiv& \varphi_{n_1 l_1 j_1 m_1 m_{t1}} (\vec{r}_1 \sigma_1 \tau_1) \\ \nonumber   
   &=& \sum_{m_1 m_{s1}} (l_1 m_1 \textstyle{\frac{1}{2}} m_{s1} | j_1 m_1)\varphi_{n_1 l_1 m_1}(\vec{r}_1) \chi_{m_{s1}}(\sigma_1) \chi_{m_{t1}}(\tau_1)
\end{eqnarray}
with $m_{t1}{=}\textstyle{\frac{1}{2}} (\textstyle{-\frac{1}{2}})$ for a proton (neutron). With the abbreviations $\alpha_i\equiv n_i l_i j_i m_i m_{ti}$ and $\vec{x}_i \equiv \vec{r}_i \sigma_i \tau_i$, a two-nucleon antisymmetric basis state is obtained as
\begin{equation}\label{SDtwonucl}
\langle \vec{x}_1 \vec{x}_2 | \alpha_1 \alpha_2\rangle = \frac{1}{\sqrt{2}} (\varphi_{\alpha_1}(\vec{x}_1) \varphi_{\alpha_2}(\vec{x}_2) - \varphi_{\alpha_1}(\vec{x}_2) \varphi_{\alpha_2}(\vec{x}_1)) \; .
\end{equation}
For $A$ nucleons, this is generalized in a straightforward way using the concept of Slater determinants (SD)
\begin{eqnarray}\label{SDbasis}
\langle \vec{x}_1\ldots\vec{x}_A|\alpha_1\ldots\alpha_A \rangle = \frac{1}{\sqrt{A!}} 
\begin{vmatrix}\varphi_{\alpha_1}(\vec{x}_1) & \ldots & \varphi_{\alpha_1}(\vec{x}_A) \\
\ldots & \ldots & \ldots \\
\varphi_{\alpha_A}(\vec{x}_1) & \ldots & \varphi_{\alpha_A}(\vec{x}_A)
\end{vmatrix} \;.
\end{eqnarray}
The basis states (\ref{SDbasis}) can be conveniently represented in the second quantization with the help of anticommuting nucleon creation and annihilation operators $a^\dagger_{\alpha_i}$, $a_{\alpha_j}$, i.e.,
\begin{equation}\label{SDbasis2nd}
    a^\dagger_{\alpha_A}\ldots a^\dagger_{\alpha_1} |0\rangle \;,
\end{equation}
with the vacuum state $|0\rangle$. When employing the basis (\ref{SDbasis}) or (\ref{SDbasis2nd}), the Hamiltonian (\ref{eq:ham_Ham}) is block diagonal in parity, $\pi{=}(-1)^{\sum_i l_i}$, the total angular momentum projection $M{=}\sum_i m_i$, and the total isospin projection $M_T{=}\sum_i m_{ti}{=}\textstyle{\frac{1}{2}}(Z-N)$ where $Z$($N$) is the proton (neutron) number. Consequently, this type of basis is called the $M$-scheme basis. It is used frequently in NCSM calculations.

A connection between the Jacobi-coordinate and the SD bases is facilitated by transformation properties of the HO wave functions. Following, e.g., Ref. \cite{PhysRevC.5.1534}, the HO wave functions depending on the coordinates (\ref{jacobiam11})
transform as
\begin{eqnarray}\label{ho_tr}
&&\sum_{m_1 m_2} (l_1 m_1 l_2 m_2|{\cal Q M_Q}) \, \varphi_{n_1 l_1 m_1}(\vec{r}_1) 
\varphi_{n_2 l_2 m_2}(\vec{r}_2) 
\nonumber \\
&=&\sum_{n l m N L M} \langle nl NL {\cal Q}|n_1 l_1 n_2 l_2 {\cal Q}\rangle_{\frac{{\cal M}_2}{{\cal M}_1}} \; 
   (l m L M|{\cal Q M_Q}) \, \varphi_{nlm}(\vec{\xi}_{1}) \varphi_{NLM}(\vec{\xi}_0)
\; ,
\end{eqnarray}
where $\langle nl NL {\cal Q}|n_1 l_1 n_2 l_2 {\cal Q}\rangle_{\frac{{\cal M}_2}{{\cal M}_1}}$ 
is the general HO bracket for two particles with mass ratio $\frac{{\cal M}_2}{{\cal M}_1}$ and ${\cal Q \; M_Q}$ the total angular momentum and its projection. In the case discussed here with the coordinate transformation (\ref{jacobiam11}), we have ${\cal M}_1{=}{\cal M}_2$, i.e., the mass ratio is one.  

The NCSM trial wave function can be written as 
\begin{equation}\label{NCSM_wav}
 \ket{\Psi^{J^\pi T}_{A M_T}} = \sum_{N=0}^{N_{\rm max}}\sum_i c_{Ni M_T}^{J^\pi T}\ket{ANiJ^\pi T M_T}\; .
\end{equation}
Here, $N$ denotes the total number of HO excitations of all nucleons above the minimum configuration,  $J^\pi T$ are the total angular momentum, parity and isospin, and $i$ additional quantum numbers. The sum over $N$ is restricted by parity to either an even or odd sequence.
The basis is further characterized by the frequency $\Omega$ of the HO
well as discussed earlier. The expansion coefficients $c_{Ni M_T}^{J^\pi T}$ are determined by a diagonalization of the Hamiltonian (\ref{eq:ham_Ham}). The $N_{\rm 0min}$ of the minimum configuration ($N_{\rm max}{=}0$) is determined by the Pauli principle. For example, for $^3$H, $^3$He, $^4$He, the $N_{\rm 0min}{=}0$, for, e.g., $^6$Li, $^{10}$B, $^{13}$C, the $N_{\rm 0min}{=}2,6,9$, respectively. For a given basis state, the value of $N$ in Eq.~(\ref{NCSM_wav}) is given by $N{=}\sum_{i=1}^A (2n_i+l_i) - N_{\rm 0min}$.   

The expansion (\ref{NCSM_wav}) is typically used when using the Jacobi coordinate basis. When working in the single-particle basis, we use the $M$-scheme and the basis expansion is rather
\begin{equation}\label{NCSM_wavSD}
 \ket{\Psi^{\pi}_{A M M_T}} = \sum_{N=0}^{N_{\rm max}}\sum_i c_{NiMM_T}^{\pi}\ket{ANi\pi M M_T}\; .
\end{equation}
The expansion coefficients $c_{NiMM_T}^{\pi}$ are again determined by a diagonalization of the Hamiltonian (\ref{eq:ham_Ham}). We note that the index $i$ represents in general a different set of additional quantum numbers than that appearing in Eq.~(\ref{NCSM_wav}). The $M$-scheme basis dimension is typically very large, e.g, $10^7{-}10^9$, but the Hamiltonian matrix is sparse. It is then efficient to apply the Lanczos diagonalization algorithm~\cite{Lanczos1950} to obtain a set of the lowest eigenstates of the system. 

The HO basis  within the $N_{\rm max}$ truncation is the only possible one that allows an exact factorization of the c.m. motion for the eigenstates, even when working with single-particle coordinates and Slater determinants. Calculations performed with the two alternative coordinate choices are completely equivalent. Let's denote the eigenstates obtained from the Jacobi-coordinate trial expansion~(\ref{NCSM_wav}) $\langle \vec{\xi}_1 \ldots \vec{\xi}_{A-1} \sigma_1 \ldots \sigma_A \tau_1 \ldots \tau_A | A \lambda J^\pi M T M_T\rangle$, i.e., 
\begin{equation}\label{NCSM_eq}
\hat{H} \ket{A \lambda J^\pi T M_T} = E_{\lambda M_T} ^{J^\pi T} \ket{A \lambda J^\pi T M_T} \; ,
\end{equation}
with $\hat{H}$ given by Eq.~(\ref{eq:ham_Ham}) and $\lambda$ labeling eigenstates with identical $J^\pi T$. The Jacobi coordinates are defined by generalizing (\ref{jacobiam11}), e.g., 
\begin{equation}\label{xiAm1}
\vec{\xi}_{A-1}{=}\sqrt{\textstyle{\frac{A-1}{A}}}\left[\textstyle{\frac{1}{A-1}}\left(\vec{r}_1+\vec{r}_2 + \ldots+ \vec{r}_{A-1}\right)-\vec{r}_{A}\right] \; .
\end{equation}
The relationship between the Jacobi coordinate and the SD eigenstates obtained from the trial expansion (\ref{NCSM_wavSD}) is 
\begin{eqnarray}\label{state_relation}
&&\langle \vec{r}_1 \ldots \vec{r}_A \sigma_1 \ldots \sigma_A \tau_1 \ldots \tau_A 
  | A \lambda J^\pi M T M_T\rangle_{\rm SD} \\ \nonumber
  &=& 
\langle \vec{\xi}_1 \ldots \vec{\xi}_{A-1} \sigma_1 \ldots \sigma_A \tau_1 \ldots \tau_A 
| A \lambda J^\pi M T M_T\rangle \varphi_{000}(\vec{\xi}_0) \; ,
\end{eqnarray}
with the c.m. of the $A$-nucleon system $\vec{R}^{(A)}_{\rm c.m.}{=}\sqrt{\frac{1}{A}}\vec{\xi}_0$. In order to select the physical SD eigenstates satisfying the factorization (\ref{state_relation}), one typically applies the Lawson projection~\cite{GLOECKNER1974313} that pushes SD eigenstates with the c.m. in excited HO configurations to high energy relative to the c.m. $0\hslash\Omega$ eigenstates. This is achieved by adding a projection term $\beta (\hat{H}_{\rm c.m.}^{\rm HO}-\textstyle{\frac{3}{2}}\hslash\Omega)$ to the intrinsic Hamiltonian (\ref{eq:ham_Ham}) with $\beta$ a positive parameter (typically $\beta{\sim}1{-}10$). The physical eigenvalues and eigenvectors are independent of $\beta$. The $\hat{H}_{\rm c.m.}^{\rm HO}$ is the c.m. HO Hamiltonian of the $A$-nucleon system.

It should be noted that in general the isospin $T$ is only approximately conserved. NCSM calculations typically fully include isospin breaking originating from the Coulomb interaction and strong force contributions present in the $\hat{V}^{\rm NN}$ (\ref{H}).
 
\begin{figure}
    \centering
    \includegraphics[width=0.8\textwidth]{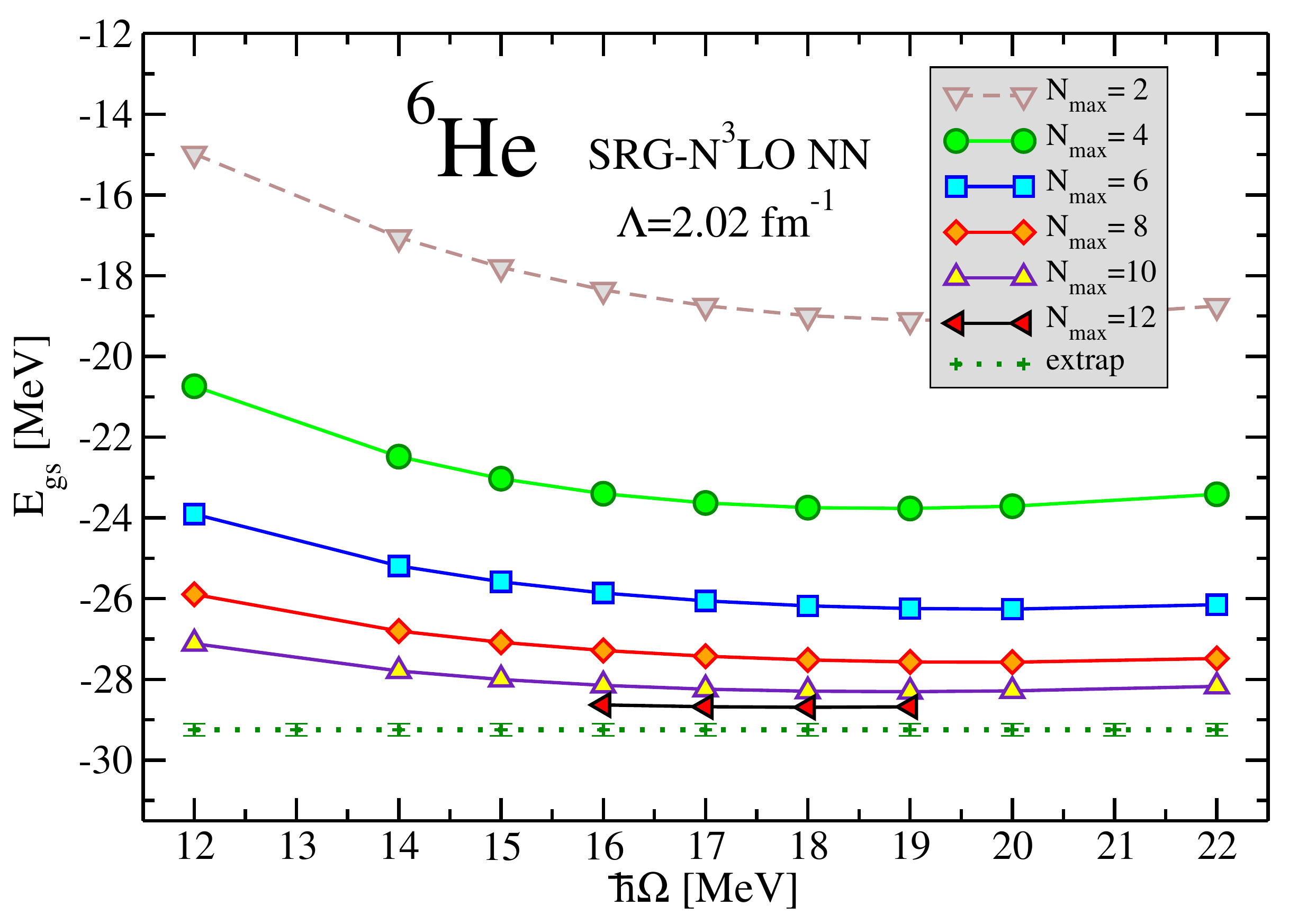}
    \caption{Ground-state energy of $^6$He calculated within the NCSM using the SRG-evolved N$^3$LO NN potential from Ref.~\cite{Entem2003} with $\Lambda{=}2.02$ fm$^{-1}$. The dependence on the HO frequency for different $N_{\rm max}$ basis sizes is shown. The points with error bars represent the results of the exponential extrapolation. Adapted from Ref.~\cite{Baroni2013}.}
    \label{fig:6He_gs}
\end{figure}
 
Convergence of the HO expansion with increasing $N_{\rm max}$ values can be accelerated by the use of unitary SRG transformations in momentum space~\cite{Bogner2007,PhysRevC.77.064003,Bogner201094,Jurgenson2009,Jurgenson2011} discussed in more detail in Section Similarity Renormalization Group Transformations. The NCSM calculations are variational. This is demonstrated in Fig.~\ref{fig:6He_gs} showing the ground-state (g.s.) energy convergence of $^6$He using an SRG renormalized chiral NN interaction. Calculations for a range of HO frequencies and increasing basis sizes characterized by $N_{\rm max}$ are presented. Clearly, with increasing $N_{\rm max}$ the ground state energy decreases and the dependence on the variational parameter $\hslash\Omega$ decreases. It is feasible to perform an extrapolation to infinite basis using the exponential function $E (N_{\rm max}) = E_{\infty} + a e^{- b N_{\rm max}}$ and vary the number of $N_{\rm max}$ points and the HO frequency to estimate its uncertainty. 

When the SRG renormalization of the Hamiltonian is applied, the many-body wave functions obtained in the NCSM (as well as in the NCSM/RGM and the NCSMC) will be in general renormalized as well. This fact is in particular important to keep in mind when using low SRG parameters, i.e., $\Lambda\lesssim 2$ fm$^{-1}$. Finally, we note that with the HO basis sizes typically used ($N_{\rm max}{\sim}10{-}14$), the NCSM $\ket{A \lambda J^\pi T}$ eigenstates lack correct asymptotic behavior for weakly-bound states and always have incorrect asymptotic behavior for resonances.

\section{\textit{Ab Initio} Approach to Nuclear Reactions}
\subsection{Binary-Cluster Resonating-Group Method}
\label{RGM}
The RGM is a microscopic cluster technique in which the $A$-nucleon Hilbert space is spanned by wave functions describing a system of two or more clusters in relative motion. For the two clusters the channel (basis) states of total angular momentum $J$, parity $\pi$, and isospin $T$ are given by:
\begin{eqnarray}
|\Phi^{J^\pi T}_{\nu r}\rangle &=& \Big [ \big ( \left|A{-}a\, \alpha_1 I_1^{\,\pi_1} T_1\right\rangle \left |a\,\alpha_2 I_2^{\,\pi_2} T_2\right\rangle\big ) ^{(s T)}\,Y_{\ell}\left(\theta_{A-a,a},\phi_{A-a,a}\right)\Big ]^{(J^\pi T)}
\nonumber \\
&\times& \,\frac{\delta(r-r_{A-a,a})}{rr_{A-a,a}}\,.\label{eq:rgm-basis}
\end{eqnarray}
To simplify the notation, here we use parenthesis and square brackets to indicate implicitly the angular momentum couplings and the overall coupling to $J^\pi$ and T, respectively, and denote the intrinsic (antisymmetric) wave functions of the first and second clusters containing $A{-}a$ and $a$ nucleons ($a{<}A$) by the kets $ \left|A{-}a\, \alpha_1 I_1^{\,\pi_1} T_1\right\rangle$ and $\left |a\,\alpha_2 I_2^{\,\pi_2} T_2\right\rangle$, respectively. These are characterized by angular momentum (iso-spin) quantum numbers $I_1$ and $I_2$ ($T_1$ and $T_2$) coupled together to form channel spin $s$ (total isospin $T$). For the clusters' parity and additional quantum numbers we use, respectively, the notations $\pi_i$, and $\alpha_i$, with $i=1,2$. The clusters' centers of mass are separated by the relative coordinate 
\begin{equation}
\vec r_{A-a,a} = (r_{A-a,a},\theta_{A-a,a},\phi_{A-a,a})= \frac{1}{A - a}\sum_{i = 1}^{A - a} \vec r_i - \frac{1}{a}\sum_{j = A - a + 1}^{A} \vec r_j\,,
\label{eq:rgm-relative}
\end{equation}
where $\{\vec{r}_i, i=1,2,\cdots,A\}$ are as usual the $A$ single-particle coordinates, and the channel states~(\ref{eq:rgm-basis}) describe partial waves of relative motion characterized by the orbital angular momentum $\ell$. For brevity, we will group all relevant quantum numbers into a cumulative index $\nu=\{A{-}a\,\alpha_1I_1^{\,\pi_1} T_1;\, a\, \alpha_2 I_2^{\,\pi_2} T_2;$ $\, s\ell\}$.
 
The channel states (\ref{eq:rgm-basis}) can be used as a basis to represent the $A$-nucleon wave function according to
\begin{equation}
|\Psi^{J^\pi T}\rangle = \sum_{\nu} \int dr \,r^2\frac{\gamma^{J^\pi T}_\nu(r)}{r}\,\hat{\mathcal A}_{\nu}\,|\Phi^{J^\pi T}_{\nu r}\rangle\,, \label{trial}
\end{equation}
where $\gamma^{J^\pi T}_\nu(r)$ are continuous (variational) amplitudes of relative motion between the pairs of clusters. 
However, to preserve the Pauli principle one has to first introduce the appropriate inter-cluster antisymmetrizer, which is schematically given by
\begin{equation}
\hat{\mathcal A}_{\nu}=\sqrt{\frac{(A-a)!a!}{A!}}\left( 1+\sum_{P\neq id}(-)^pP\right)\,,
\label{antisym}
\end{equation}   
where the sum runs over all possible permutations of nucleons $P$ (different from the identical one) 
%$P$ are permutations
that can be carried out between two different clusters (of $A-a$ and $a$ nucleons, respectively), and $p$ is the number of interchanges characterizing them. The operator~(\ref{antisym}) is labeled by the channel index $\nu$ to signify that its form depends on the mass partition, $(A-a,a)$, of the channel state to which it is applied. 
Indeed, while the clusters' wave functions are antisymmetric, the basis states~(\ref{eq:rgm-basis}) are not antisymmetric under exchange of nucleons belonging to different clusters. It should be noted that each permutation in $\hat{\mathcal A}_{\nu}$ changes the relative coordinate between the clusters, as a result of the rearrangement of the nucleons. This complication is circumvented by the introduction of the Dirac's delta function of Eq.~(\ref{eq:rgm-basis}), thanks to which the inter-cluster antisymmetrizer does not affect the amplitudes $\gamma^{J^\pi T}_\nu(r)$ that depend on the integration variable $r$.

By using the ansatz of eq.~(\ref{trial}) for the wave function and projecting (from the left) on the basis states~(\ref{eq:rgm-basis}), the $A$-nucleon Schr\"odinger equation %~(\ref{eq:SE}) 
is mapped into a one-dimensional coupled-channel equation for the unknown amplitudes of relative motion $\gamma^{J^\pi T}_\nu(r)$
\begin{equation}
\sum_{\nu}\int dr \,r^2\left[{\mathcal H}^{J^\pi T}_{\nu^\prime\nu}(r^\prime, r)-E\,{\mathcal N}^{J^\pi T}_{\nu^\prime\nu}(r^\prime,r)\right] \frac{\gamma^{J^\pi T}_\nu(r)}{r} = 0\,,\label{RGMeq}
\end{equation}
where $E$ denotes the total energy in the c.m.\ frame. The two integration `kernels' in the above equation are the Hamiltonian kernel,
\begin{equation}
{\mathcal H}^{J^\pi T}_{\nu^\prime\nu}(r^\prime, r) = \left\langle\Phi^{J^\pi T}_{\nu^\prime r^\prime}\right|\hat{\mathcal A}_{\nu^\prime}\hat{H}\hat{\mathcal A}_{\nu}\left|\Phi^{J^\pi T}_{\nu r}\right\rangle\,,\label{H-kernel}
\end {equation}
and the norm kernel,
\begin{equation}
{\mathcal N}^{J^\pi T}_{\nu^\prime\nu}(r^\prime, r) = \left\langle\Phi^{J^\pi T}_{\nu^\prime r^\prime}\right|\hat{\mathcal A}_{\nu^\prime}\hat{\mathcal A}_{\nu}\left|\Phi^{J^\pi T}_{\nu r}\right\rangle\,,\label{N-kernel}
\end{equation}
which are nothing else than the matrix elements of the Hamiltonian and identity operators over the antisymmetric basis states $\hat{\mathcal A}_{\nu}\left|\Phi^{J^\pi T}_{\nu r}\right\rangle$, and contain all the nuclear structure and antisymmetrization properties of the problem. The somewhat unusual presence of a norm kernel is the result of the non-orthogonality of the antisymmetric basis states owing to the operator $\hat{\mathcal A}_{\nu}$. The exchange terms of this antisymmetrization operator are also responsible for  the non-locality of the two kernels, which couple amplitudes at different distances $r$ and $r^\prime$.

In Eq.~(\ref{H-kernel}),  $\hat{H}$ is the microscopic $A-$nucleon Hamiltonian of Eq.~(\ref{H}), which can be conveniently separated into the intrinsic Hamiltonians for the $(A-a)$- and $a$-nucleon systems, respectively $\hat{H}_{(A-a)}$ and $\hat{H}_{(a)}$, plus the relative motion Hamiltonian
\begin{equation}\label{eq:Hrgm}
\hat{H}=T_{\rm rel}(r)+{V}_{\rm C}(r)+{\mathcal V}_{\rm rel} + \hat{H}_{(A-a)}+ \hat{H}_{(a)}\,.
\end{equation}
Here, $T_{\rm rel}(r)$ is the relative kinetic energy between the two clusters, ${V}_{\rm C}(r)=Z_{1\nu}Z_{2\nu}e^2/r$ ($Z_{1\nu}$ and $Z_{2\nu}$ being the charge numbers of the clusters in channel $\nu$) the average Coulomb interaction between pairs of clusters, and ${\mathcal V}_{\rm rel}$ is a localized relative (inter-cluster) potential given by:
\begin{align}
{\mathcal V}_{\rm rel} &= \sum_{i=1}^{A-a}\sum_{j=A-a+1}^A \hat{V}^{NN}_{ij} + \sum_{i<j=1}^{A-a}\sum_{k=A-a+1}^A \hat{V}^{3N}_{ijk} + \sum_{i=1}^{A-a}\sum_{j<k=A-a+1}^A \hat{V}^{3N}_{ijk} - {\bar{V}}_{\rm C}(r)\label{pot}\,.
\end{align}
Besides the nuclear components of the interactions between nucleons belonging to different clusters, it is important to notice that the overall contribution to the relative potential~(\ref{pot}) coming from the Coulomb interaction,
\begin{equation}
\label{locCoul}
\sum_{i=1}^{A-a}\sum_{j=A-a+1}^A\left(\frac{e^2(1+\tau^z_i)(1+\tau^z_j)}{4|\vec r_i-\vec r_j|} -\frac{1}{(A-a)a}{\bar V}_{\rm C}(r)\right)\,,
\end{equation}
 is also localized, presenting an $r^{-2}$ behavior, as the distance $r$ between the two clusters increases.

Using the expression of Eq.~(\ref{eq:Hrgm}) for the nuclear Hamiltonian and Eqs.~(\ref{eq:rgm-basis}) and (\ref{antisym}), it is possible to demonstrate that the norm and Hamiltonian kernels can be separated into ``full-space" and ``localized" components as
\begin{equation}
{\mathcal N}^{J^\pi T}_{\nu^\prime\nu}(r^\prime, r)
= \delta_{\nu^\prime\nu}\frac{\delta(r^\prime-r)}{r^\prime r} + {\mathcal N}^{\rm ex}_{\nu^\prime\nu}(r^\prime, r) \,,%\sqrt{ \frac{ (A-a)!a! }{ (A-a^\prime)! a^\prime!} }
\label{N-kernel-2}
\end{equation}
and
\begin{equation}
{\mathcal H}^{J^\pi T}_{\nu^\prime\nu}(r^\prime, r) = \left[{ T}_{\rm rel}(r')+{\bar{V}}_C(r')+E_{\alpha_1'}^{I_1'T_1'} +E_{\alpha_2'}^{I_2'T_2'}\right]
\mathcal{N}_{\nu'\nu}^{J^\pi T}(r', r)+\mathcal{V}^{J^\pi T}_{\nu' \nu}(r',r)\,,
\label{H-kernel-2}
\end {equation}
where $E_{\alpha_i}^{I_i^{\pi_i} T_i}$ are the energy eigenvalues of the $i$-th cluster ($i=1,2$).
The exchange part of the norm, ${\mathcal N}^{\rm ex}_{\nu^\prime\nu}(r^\prime, r)$, and the potential kernel, $\mathcal{V}^{J^\pi T}_{\nu' \nu}(r',r)$, given by
\begin{equation}
{\mathcal N}^{\rm ex}_{\nu^\prime\nu}(r^\prime, r) = 
\left\{
\begin{array}{ll}
	\left\langle\Phi^{J^\pi T}_{\nu^\prime r^\prime}\right| \sum_{P\neq id}(-)^p \hat P \left|\Phi^{J^\pi T}_{\nu r}\right\rangle& \quad {\rm if}~a^\prime=a\\
	\\
%	\sqrt{ \tfrac{ (A-a)!a! }{ (A-a^\prime)! a^\prime!} }\sum_{n^\prime n}R_{n^\prime\ell^\prime}(r^\prime)R_{n\ell}(r)\left\langle\Phi^{J^\pi T}_{\nu^\prime n^\prime}\right| 1-\sum_{P}(-)^p P \left|\Phi^{J^\pi T}_{\nu n}\right\rangle& \quad {\rm if}~a^\prime\neq a
	\left\langle\Phi^{J^\pi T}_{\nu^\prime r^\prime}\right| \sqrt{\frac{ A! }{ (A-a^\prime)! a^\prime !}} \hat {\mathcal A}_\nu \left|\Phi^{J^\pi T}_{\nu r}\right\rangle& \quad {\rm if}~a^\prime\neq a
\end{array}
\right .
%\left\langle\Phi^{J^\pi T}_{\nu^\prime n^\prime}\right| \sqrt{\tfrac{ A! }{ (A-a^\prime)! a^\prime !}} \hat {\mathcal A}_\nu - \delta_{a^\prime a}\left|\Phi^{J^\pi T}_{\nu n}\right\rangle
\label{Nex-kernel}
\end{equation}
and 
\begin{equation}
\mathcal{V}^{J^\pi T}_{\nu' \nu}(r',r) = 
	\left\langle\Phi^{J^\pi T}_{\nu^\prime r^\prime}\right| 
%	%\tfrac 12 \left(\hat{\mathcal A}_{\nu^\prime} {\mathcal V}_{\rm rel}  \sqrt{\tfrac{ A! }{ (A-a)! a!}} + 
	\sqrt{\frac{ A!}{ (A-a^\prime)! a^\prime !}} {\mathcal V}_{\rm rel}\hat{\mathcal A}_{\nu} 
	% \right)
	\left|\Phi^{J^\pi T}_{\nu r}\right\rangle\,,
	\label{V-kernel}
\end{equation}
are both localized quantities because of the short range of the non-orthogonality induced by the non-identical permutations and of the nuclear interaction, and can be computed by expanding the channel states~(\ref{eq:rgm-basis}) on a complete basis of localized states. While the calculation of these kernels constitutes the main challenge in the application of the RGM approach, it can be accomplished using the same type of powerful and well-established techniques used for the solution of the $A$-nucleon problem for bound states, discussed in Section No-Core Shell Model. In particular, the RGM can be implemented using NCSM wave functions for the clusters in a formalism known as NCSM/RGM~\cite{Quaglioni2008,Quaglioni2009,Hupin2013}. A recent review of this method can be found in ref.~\cite{Navratil2016}. 
\begin{figure}[t]
\centering
\includegraphics[width=\textwidth]{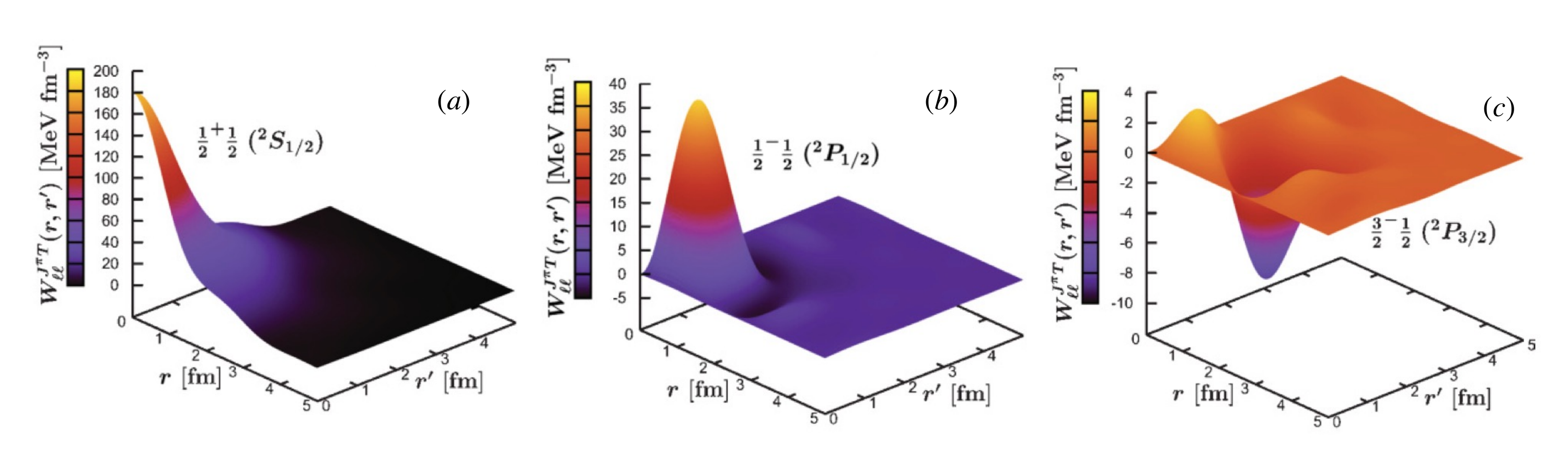}
\caption{An example of the non-local potential $W^{J^\pi T}_{\nu^\prime \nu}(r^\prime,r)$ for the $n$+$^4$He (or $n+\alpha$) system obtained within the NCSM/RGM approach. The three panels show respectively the $n+^4$He potential for the (a) repulsive $S$-wave and for the $P$-waves in the (b) $J^\pi=1/2^-$ and (c) $J^\pi=3/2^-$ channels. The most attractive potential surface is found in the $J^\pi=3/2^-$ channel, corresponding to the g.s.\ of the $^5$He resonance (see also Section DT and D$^3$He Fusion).}
\label{fig:W}
\end{figure}

An important point to notice, is that Eq.~(\ref{RGMeq}) does not represent a system of multichannel Schr\"odinger equations, and $\gamma^{J^\pi T}_\nu(r)$ do not represent Schr\"odinger wave functions. This feature, due to the presence of the norm kernel ${\mathcal N}^{J^\pi T}_{\nu^\prime\nu}(r^\prime, r)$,  can be corrected by introducing normalized Schr\"odinger wave functions
\begin{equation}
\frac{\chi^{J^\pi T}_\nu(r)}{r} = \sum_{\gamma}\int dy\, y^2 {\mathcal N}^{\frac12}_{\nu\gamma}(r,y)\,\frac{g^{J^\pi T}_\gamma(y)}{y}\,,
\label{eq:rgm-ortho}
\end{equation}
where ${\mathcal N}^{\frac12}$ is the square root of the norm kernel, and applying the inverse-square root of the norm kernel, ${\mathcal N}^{-\frac12}$, to both left and right-hand side of the square brackets in Eq.~(\ref{RGMeq}).  By means of this procedure, known as orthogonalization and explained in more detail in ref.~\cite{Quaglioni2009}, one obtains a system of multichannel Schr\"odinger equations:
\begin{eqnarray}
&&\left\{-\frac{\hslash^2}{2\mu_{\nu}r^2} \frac{d}{dr} \left(r^2\frac{d}{dr}\right ) + \frac{\hslash^2 \ell(\ell+1)}{2\mu_\nu r^2} +  \bar V_{\rm C}(r) \right\} \frac{\chi^{J^\pi T}_{\nu} (r)}{r} \nonumber \\
&+& \sum_{\nu^\prime}\int dr^\prime\,r^{\prime\,2} \,W^{J^\pi T}_{\nu \nu^\prime}(r,r^\prime)\,\frac{\chi^{J^\pi T}_{\nu^\prime}(r^\prime)}{r^\prime} = (E - E_\nu) \frac{\chi^{J^\pi T}_{\nu} (r)}{r}\,,
\label{eq:1DrgmSE}
\end{eqnarray} 
where $\mu_\nu$ is the reduced mass of the clusters in channel $\nu$, $E_{\nu}=E_{\alpha_1}^{I_1^{\pi_1} T_1} + E_{\alpha_2}^{I_2^{\pi_2} T_2}$ is the threshold at which the reaction channels become open and 
%\begin{eqnarray}
%[ T_{\rm rel}(r) + \bar V_{\rm C}(r) -(E - E_{\alpha_1}^{I_1^{\pi_1} T_1} - E_{\alpha_2}^{I_2^{\pi_2} T_2})]\frac{\chi^{J^\pi T}_{\nu} (r)}{r} + \sum_{\nu^\prime}\int dr^\prime\,r^{\prime\,2} \,W^{J^\pi T}_{\nu \nu^\prime}(r,r^\prime)\,\frac{\chi^{J^\pi T}_{\nu^\prime}(r^\prime)}{r^\prime} = 0,\label{r-matrix-eq}
%\end{eqnarray} 
$W^{J^\pi T}_{\nu^\prime \nu}(r^\prime,r)$ are the non-local potentials between the two clusters, which depend upon the channel of relative motion but do not depend upon the energy $E$ of the system. An example of such potentials is shown in fig.~\ref{fig:W}.

With the exception of the presence of a non-local interaction, Eq.~(\ref{eq:1DrgmSE}) resembles closely the radial Schr\"odinger equation for a two-nucleon system and can be easily solved (subject to appropriate boundary conditions) using the same numerical methods to obtain bound-state wave functions and binding energies or scattering wave functions and  scattering matrix, from which any other scattering and reaction observable can be calculated.

Finally, we note that while in principle the summation in the expansion~({\ref{trial}}) runs over all possible partitions of the $A$-nucleon systems into two bound clusters of nucleons (including states in which the clusters are in a bound excited state), in practical calculations for computational reasons it is restricted to just a few channels, typically those that are energetically open. While such an approximation is energetically justified, it constitutes the main limitation of the RGM in recovering the full $A$-nucleon dynamics at short distances, where single-particle many-body correlations play an important role.

\subsection{Three-Cluster RGM}
\begin{figure}[t]
\centering
\includegraphics[width=4cm,clip=,draft=false, angle=0]{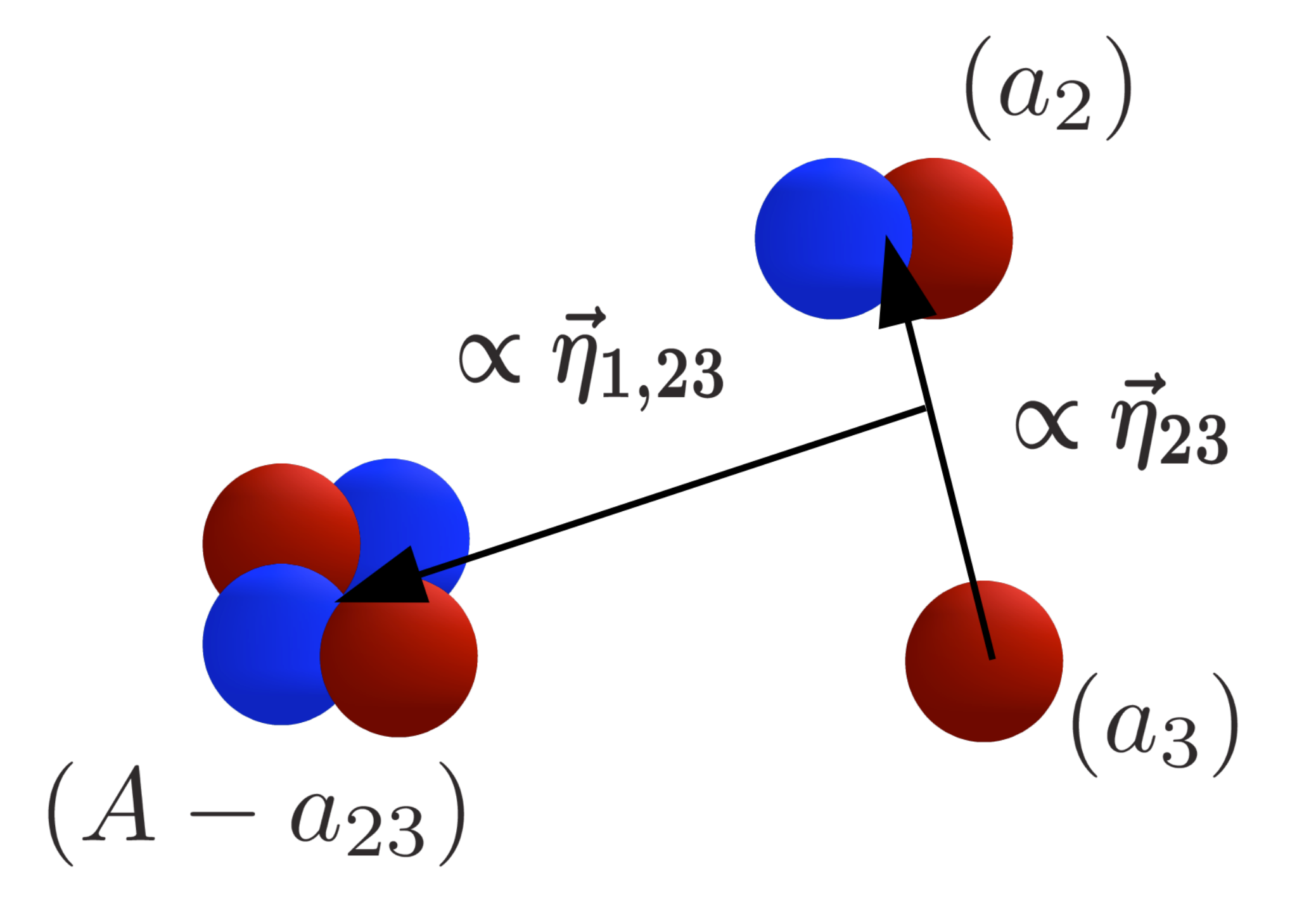}
\caption{Example of Jacobi relative coordinates for three-cluster configurations.}
\label{FigCoor}
\end{figure}
The RGM can also be extended to the description of three and in principle several clusters in relative motion with respect to each other.
To accomplish this, it is convenient to introduce a suitable set of Jacobi relative coordinates among the clusters. For example, for a system of three clusters respectively of mass number $A-a_{23}$, $a_2$, 
and $a_3$ ($a_{23}=a_2+a_3< A$) one can choose 
\begin{equation}
        \vec\eta_{1,23}  = (\eta_{1,23}, \theta_{{\eta}_{1,23}}, \phi_{{\eta}_{1,23}})  \label{eq:etay}
         = \sqrt{\frac{a_{23}}{A(A-a_{23})}}  \sum_{i=1}^{A-a_{23}} \vec{r}_i - \sqrt{\frac{A-a_{23}}{A\,a_{23}}} \sum_{j=A-a_{23}+1}^A \vec{r}_j \,,
\end{equation}
the relative vector proportional to the displacement between the center of mass of the first cluster and that of the residual two fragments, and
\begin{equation}
        \vec\eta_{23}  = (\eta_{23}, \theta_{{\eta}_{23}},\phi_{{\eta}_{23}}) \label{eq:etax} 
         =\sqrt{\frac{a_3}{a_{23}\,a_2}}  \sum_{i=A-a_{23}+1}^{A-a_3} \vec{r}_i - \sqrt{\frac{a_2}{a_{23}\,a_3}} \sum_{j=A-a_3+1}^A \vec{r}_j\,,      
\end{equation}
the relative coordinate proportional to the distance between the centers of mass of cluster
2 and 3 (See figure \ref{FigCoor}). Here, $\vec{r}_i$ denotes the position vector of the $i$-th nucleon.
Introducing the associated hyperspherical harmonics (HH) coordinates $\rho_\eta$ and $\alpha_\eta$
\begin{eqnarray}
        \eta_{23} &=& \rho_\eta \sin\alpha_\eta\, \\
        \eta_{1,23} &=& \rho_\eta \cos\alpha_\eta\,, 
\end{eqnarray}
one can then define RGM channel states for three clusters as 
\begin{eqnarray}
	|\Phi^{J^\pi T}_{\nu K \rho} \rangle &=& 
	\Big[\Big(|A-a_{23}~\alpha_1I_1^{\pi_1}T_1\rangle 
	\left (|a_2\, \alpha_2 I_2^{\pi_2} T_2\rangle |a_3\, \alpha_3 I_3^{\pi_3}T_3\rangle \right)^{(s_{23}T_{23})}\Big)^{(ST)} \nonumber \\
	&\times& {\mathcal Y}_{L}^{K\ell_x\ell_y}(\Omega_\eta)\Big]^{(J^{\pi}T)} 
\;\frac{\delta(\rho - \rho_\eta)}{\rho^{5/2}\,\rho_\eta^{5/2}}\,,
	\label{eq:3bchannelHH}	
\end{eqnarray}
with $\Omega_\eta = (\alpha_\eta, \theta_{{\eta}_{1,23}}, \phi_{{\eta}_{1,23}}, \theta_{{\eta}_{23}},\phi_{{\eta}_{23}})$ and 
${\mathcal Y}_{L}^{K\ell_x\ell_y}(\Omega_\eta)$ the HH basis elements. %defined in Eq. (\ref{HHbasis}). 
The rest of the notation is similar to that introduced in eq.~(\ref{eq:rgm-basis}).
The $A$-nucleon wave function can be written as 
\begin{equation}
	|\Psi^{J^\pi T}\rangle = \sum_{\nu K} \int d\rho\, \rho^5 \frac{\gamma_{K\nu}^{J^\pi T}(\rho)}{\rho^{5/2}} \hat{\mathcal A}_\nu |\Phi^{J^\pi T}_{\nu K \rho} \rangle\,,
\end{equation}
where $\hat{\mathcal A}_\nu$ is an appropriate inter-cluster antisymmetrizer and the unknown hyperradial amplitudes $\gamma_{K\nu}^{J^\pi T}(\rho)$ are obtained from the solution of the one-dimensional coupled-channel equation
\begin{equation}
        \sum_{K\nu}\int d\rho \rho^5 \left[{\cal H}_{\nu'\nu}^{K'K}(\rho',\rho) - E {\cal N}_{\nu'\nu}^{K'K}(\rho',\rho) \right]\frac{\gamma^{J^{\pi}T}_{K\nu}(\rho)}{\rho^{5/2}} 
        = 0\,,
        \label{RGMrho}
\end{equation}
which is the three-cluster analog of eq.\ (\ref{RGMeq}). Also analogous, though more involved, is the formalism for computing the three-cluster Hamiltonian and norm kernels. For a detailed discussion of the three-cluster RGM formalism we refer the interested reader to, e.g., Ref.~\cite{Quaglioni2013}.

In summary, within the RGM the $A$-nucleon problem of few clusters in mutual relative motion is mapped into a more approachable problem of few point-like cluster degrees of freedom interacting through non-local potentials. Much of the complexity of the problem is transferred into the determination of the non-local interactions among the clusters, which absorb all the information about the nuclear structure and antisymmetrization properties of the system. While challenging to compute, such interactions can be obtained starting from the {\it ab initio} wave functions describing the quantum states of the nuclear clusters using a technique for the solution of the $A$-nucleon bound-state problem such as the NCSM.

\subsection{Unified Description of Bound and Continuum States: the No-Core Shell Model with Continuum}\label{subsec:ncsmc}
Achieving a comprehensive and simultaneous treatment of short-range many-body correlations and long-range few-cluster dynamics in atomic nuclei constitutes an enormous challenge. 
{\it Ab initio} bound-state techniques (which are at least in part based on the methodologies discussed in Section No-Core Shell Model) successfully describe the interior of the nuclear wave function, but are unable
to fully account for its asymptotic behavior, which in general can be dominated by the onset of clustering. 
At the same time, microscopic cluster techniques such as the RGM (discussed in Subsection Binary-Cluster Resonating-Group Method) naturally explain the asymptotic configurations of the nuclear wave function, but tend to underestimate many-body correlations. 

A solution to this conundrum is provided by generalized cluster expansions, where  the translational-invariant ansatz for the $A$-nucleon wave function is given by a superposition of discrete (and localized) energy eigenstates of the aggregate $A$-nucleon system, $\ket{A \lambda J^\pi T}$ (see Section No-Core Shell Model), and continuous RGM binary-cluster (and/or multi-cluster, depending on the particle-emission channels characterizing the nucleus in consideration) channel states, $\hat{\mathcal{A}}_\nu\ket{\Phi_{\nu r}^{J^\pi T}}$ (see Subsection Binary-Cluster Resonating-Group Method),  according to: 
\begin{equation}
\label{NCSMC_wav}
\ket{\Psi^{J^\pi T}_A}  =  \sum_\lambda c^{J^\pi T}_\lambda \ket{A \lambda J^\pi T}  + \sum_{\nu} \int dr \, r^2 \;
                               \frac{\gamma_{\nu}^{J^\pi T}(r)}{r}
                               \;\hat{\mathcal{A}}_\nu\ket{\Phi_{\nu r}^{J^\pi T}}.
\end{equation}
The unknown discrete  $c^{J^\pi T}_\lambda$ and continuous $\gamma_\nu^{J^\pi T}(r)$ %${\mathcal G}_{\nu}^{J^\pi T}(\{{\mathcal R}\})$,
variational amplitudes can be simultaneously obtained by solving the set of coupled equations (written here in a schematic way),
\begin{equation}
\left(
\begin{array}{cc}
        {\cal E} & \bar{h}\\  
	   \bar{h}  &\overline{\mathcal{H}} 
\end{array}
\right)
\left(
\begin{array}{c}
	c \\ \chi
\end{array}
\right)  =   E \left(
\begin{array}{cc}
        {\delta}&\bar{g} \\  
        \bar{g}  &{\Delta} 
\end{array}
\right) \left(
\begin{array}{c}
c\\
\chi
\end{array}
\right)\,,
\label{eq:NCSMC-eq}
\end{equation}
where $ \chi^{J^\pi T}(r)$ are the relative motion wave functions of eq.~(\ref{eq:rgm-ortho}). The two by two block-matrices on the left- and right-hand side of Eq.~(\ref{eq:NCSMC-eq}) represent, respectively, the Hamiltonian and norm 
kernels of the generalized cluster expansion.  The upper diagonal blocks are given by the Hamiltonian (norm) matrix elements over the basis states $ \ket{A \lambda J^\pi T}$. In particular, as the basis states are eigenstates of the $A$-nucleon Hamiltonian, these are trivially given by the diagonal matrix ${\cal E}_{\lambda\lambda^\prime} = E_\lambda \delta_{\lambda\lambda^\prime}$ of the eigenenergies (the identity matrix $\delta_{\lambda\lambda^\prime}$). Similarly, those over the %orthonormalized 
continuous portion of the basis appear in the lower diagonal blocks and are given by the orthogonalized Hamiltonian kernel of the RGM, $\overline{\mathcal{H}}{=}{\mathcal N}^{-1/2}\mathcal{H}{\mathcal N}^{-1/2}$ [see also, e.g.,\ the left-hand side of Eq.~(\ref{eq:1DrgmSE})], and 
%\begin{align}
${\Delta}_{\nu\nu^\prime}(r,r^\prime) = \delta_{\nu\nu^\prime}\delta(r-r^\prime)/r r^\prime$.
%\end{align} 
The off-diagonal blocks contain the couplings between the two sectors of the basis, with 
\begin{equation}
\bar{g}_{\lambda \nu}(r)= \sum_{\nu^\prime}\int dr^\prime\, r^{\prime\,2} \bra{A\, \lambda J^\pi T} \hat{{\mathcal A}}_{\nu^\prime}\ket{\Phi^{J^\pi T}_{\nu^\prime r^\prime}} \, {\mathcal N}^{-1/2}_{\nu^\prime\nu}(r^\prime,r)
\label{g-bar}
\end{equation}
the cluster form factor, and the coupling form factor analogously given by 
\begin{equation}
\bar{h}_{\lambda \nu}(r)= \sum_{\nu^\prime}\int dr^\prime\, r^{\prime\,2} \bra{A\, \lambda J^\pi T} \hat H \hat{{\mathcal A}}_{\nu^\prime}\ket{\Phi^{J^\pi T}_{\nu^\prime r^\prime}} \, {\mathcal N}^{-1/2}_{\nu^\prime\nu}(r^\prime,r)\,.
\end{equation}
When the aggregate and clusters' wave functions are eigenstates of their respective intrinsic Hamiltonians computed within the NCSM [see Eqs.~(\ref{NCSM_wav}) and (\ref{NCSM_eq})], this approach is known as no-core shell model with continuum (NCSMC)~\cite{Baroni2013L,Baroni2013} and both kernels and form factors can be computed using the second-quantization formalism. The interested reader can find their algebraic expressions in refs.~\cite{Baroni2013L,Baroni2013}. Similar to the RGM in Subsection Binary-Cluster Resonating-Group Method, the formalism presented here can also be generalized for the description of three-cluster dynamics. A detailed presentation of such formalism in the context of the NCSMC can be found in ref.~\cite{Quaglioni2018}.

Upon orthogonalization, Eq. (\ref{eq:NCSMC-eq}) can be turned into a set of coupled-channel equations with non-local interaction terms much alike those shown in Eq.~(\ref{eq:1DrgmSE}). That is, the $A$-nucleon problem is once again mapped into a much simpler to solve few-body problem of interacting point-like clusters. A detailed description  of the orthogonalization procedure goes beyond the scope of this work and can be found in Refs.~\cite{Baroni2013} and \cite{Navratil2016}.

\subsection{\label{subsec:rmatrix}R-matrix Method}
\begin{figure}[t]
\centering
\includegraphics[width=9cm,clip=,draft=false, angle=0]{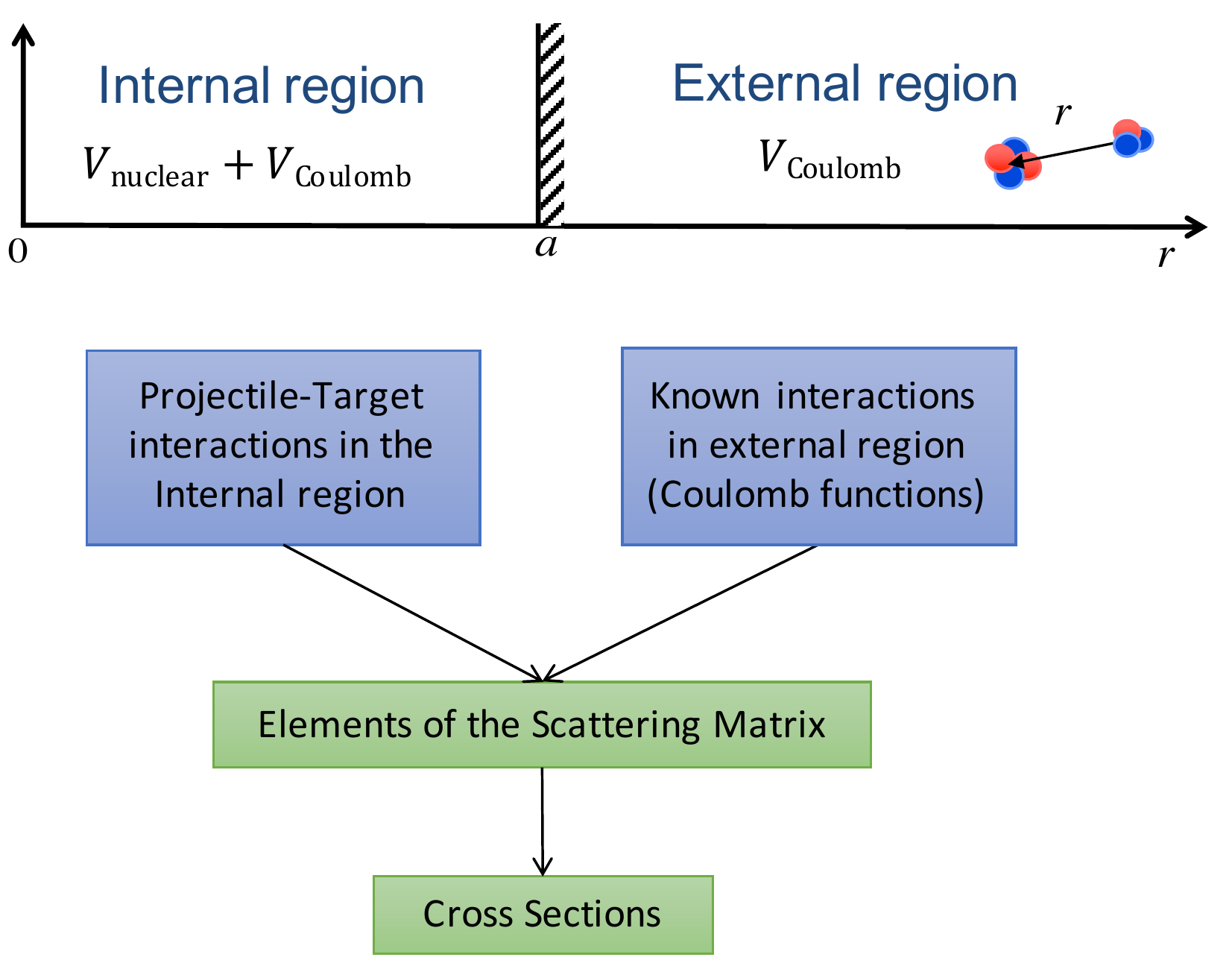}
\caption{A schematic representation of the $R$-matrix method.}
\label{fig:rmatrix}
\end{figure}

The solution of the integral-differential equations (\ref{eq:1DrgmSE}) and (\ref{RGMrho}) for the relative-motion wave function in the RGM [and simultaneously for the coefficients of the expansion on the discrete basis states in the case of Eq.~(\ref{eq:NCSMC-eq})] can be computed for both bound and continuum states working within the coupled-channel $R$-matrix method %on a Lagrange mesh 
\cite{Hesse1998,Hesse2002}.  Here we outline the formalism for the case of two clusters, a projectile and a target. The formulation for ternary clusters is analogous and can be found, e.g., in Ref.~\cite{Descouvemont:2003ys,Descouvemont:2005rc}.

In the $R$-matrix method the configuration space is separated into two regions delimited by the radius $r=a$, which is called the `matching' radius for reasons that will become clear in the following. In the internal region, the projectile and target experience both the mutual attraction owing to the nuclear force and the repulsion due to the Coulomb interaction. Here the wave function of the $A$-nucleon system can be described using a microscopic theory of nuclear forces and nuclear structure, such as the RGM or the NCSMC (see Subsections Binary-Cluster Resonating-Group Method and Unified Description of Bound and Continuum States: the No-Core Shell Model with Continuum). Because of the short-range of the nuclear force, the matching radius $a$ can be chosen large enough that in the external region the only interaction between projectile and target is the average Coulomb potential (see Subsection Binary-Cluster Resonating-Group Method) between the two. Therefore, in this region the radial wave function can be written in terms of the known solutions of the two-body problem with a Coulomb potential. For bound states, these are the 
Whittaker functions $W_{\ell}(\eta_{\nu},\kappa_{\nu}r)$,
\begin{equation}
\chi^{J^{\pi}T}_{\nu,{\rm ext}}(r)=C^{J^{\pi}T}_{\nu}W_{\ell}(\eta_{\nu},\kappa_{\nu}r)  
\end{equation}
with $C^{J^{\pi}T}_\nu$ being the asymptotic normalization constant. For continuum states, they can be written
in terms of the incoming and outgoing Coulomb functions $H^{\pm}(\eta_{\nu},\kappa_{\nu}r)=G_{\ell}(\eta_{\nu},\kappa_{\nu}r)\mp {\rm i} F_{\ell}(\eta_{\nu},\kappa_{\nu}r)$  and the elements of 
 the scattering matrix $S^{J^{\pi}T}_{\nu i}$ as:
\begin{equation}
\chi^{J^{\pi}T}_{\nu,{\rm ext}}(r)=\frac{i}{2}v_{\nu}^{-\frac{1}{2}}\left[\delta_{\nu i}
H^-_{\ell}(\eta_{\nu},\kappa_{\nu}r)-
S^{J^{\pi}T}_{\nu i}H^+_{\ell}(\eta_{\nu},\kappa_{\nu}r)\right].  
\label{scattering}
\end{equation}
They depend on the channel-state relative angular momentum $\ell$,  wave number $\kappa_\nu=\sqrt{2\mu_\nu (E-E_\nu)}/\hslash$, the relative speed $v_\nu=\hslash k_\nu/\mu_\nu$ and the dimensionless Sommerfeld parameter $\eta_\nu=Z_1Z_2 e^2/\hslash v_\nu$. The diagonal matrix elements of the scattering matrix can be parametrized by $S^{J^{\pi}T}_{\nu \nu}{=}|S^{J^{\pi}T}_{\nu \nu}| e^{2i\delta^{J^{\pi}T}_{\nu}}$ with $\delta^{J^{\pi}T}_{\nu}$ the real diagonal phase shift. As the scattering matrix is symmetric and unitary, its eigenvalues can be expressed as $e^{2i\delta^{J^{\pi}T}_{\lambda}}$ with $\delta^{J^{\pi}T}_{\lambda}$ the real eigenphase shift. We present examples of diagonal phase shift and eigenphase shift energy dependence in the following sections.  

The scattering matrix $S^{J^\pi T}_{\nu i}$ ($i$ being the initial channel) in eq.~(\ref{scattering}) is obtained by requiring the continuity of the wave function $\chi^{J^\pi T}_\nu(r)$ and of its first derivative at the matching radius $r=a$. 
This can be conveniently achieved by inserting in both left- and right-hand sides of the radial Schr\"odinger equations [ e.g.\ eq.~(\ref{eq:1DrgmSE})] the Bloch operator (with $B_\nu$ an arbitrary constant) %-Schr\"odinger equations:
\begin{equation}
\label{bloch:binary}
\hat{\cal L}_{\nu}=\frac{\hslash}{2\mu_{\nu}}\delta(r-a)\left(\frac{d}{dr}-\frac{B_{\nu}}{r} \right)\, 
\end{equation}
which has the dual function of restoring the hermiticity of the Hamiltonian
in the internal region and enforcing a continuous derivative at the matching radius \cite{R-matrix}.
This yields the Bloch-Schr\"odinger equations
\begin{eqnarray}
&&-\frac{\hslash^2}{2\mu_\nu}\left(\frac{d^2}{dr^2} - \frac{\ell(\ell+1)}{r^2} -\frac{2\mu_\nu}{\hslash^2} \bar V_{\rm C}(r) + \kappa^2_\nu \right) \chi^{J^\pi T}_{\nu,{\rm int}} (r) \nonumber \\
&+&\sum_{\nu^\prime}\int dr^\prime\,r^{\prime} \,W^{J^\pi T}_{\nu \nu^\prime}(r,r^\prime)\,\chi^{J^\pi T}_{\nu^\prime,{\rm int}}(r^\prime)+ {\cal L}_\nu \chi^{J^\pi T}_{\nu,{\rm int}} (r)= {\cal L}_\nu \chi^{J^\pi T}_{\nu,{\rm ext}} (r)\,,
\label{bloch-schr}
\end{eqnarray} 
where the internal and the external solutions are used in the left- and right-hand members, respectively, and the equation is supplemented by the continuity condition $\chi^{J^\pi T}_{\nu,{\rm int}}(a)=\chi^{J^\pi T}_{\nu,{\rm ext}}(a)$~\cite{R-matrix}. 
%Indeed, thanks to the Dirac delta function in the Bloch operator, eq.~(\ref{bloch-schr}) supplemented by the continuity equation are equivalent to the Schr\"odinger equation~(\ref{eq:1DrgmSE}) supplemented by the continuity condition $\chi^{\prime \, J^\pi T}_{\nu,{\rm int}}(a)=\chi^{\prime \, J^\pi T}_{\nu,{\rm ext}}(a)$.

In practice, the Bloch-Schr\"odinger equations are solved by expanding the radial wave function on a convenient set of basis states.
For bound-state calculations the constants $B_{\nu}$ can be chosen as the logarithmic derivative of $\chi^{J^{\pi}T}_{\nu,ext}(r)$
evaluated in the matching radius. This choice eliminates the right hand side of Eq. (\ref{bloch-schr})  
and gives rise to an eigenvalue problem. Because $\kappa_\nu$ depends on the studied binding energy,  the determination of the bound-state energy and asymptotic normalization constant $C^{J^\pi T}_\nu$ is achieved iteratively starting from an initial guess for the value of $B_\nu$, typically $B_\nu=0$ (convergence is typically reached within a few iterations). For scattering states,
the constants $B_{\nu}$ can be chosen to be zero, and the scattering matrix is obtained by solving a set of linear equations
through the calculation of the $R$-matrix, which is the inverse of the logarithmic derivative of the wave function at $r=a$ (hence the name $R$-matrix theory). The details of this procedure can be found in Ref.~\cite{Hesse1998}.   

The internal radial wave function $\chi^{J^\pi T}_{\nu,{\rm int}} (r)$ can be conveniently expanded on a set of $N$ Lagrange functions. Due to its properties, the Lagrange basis provides straightforward expressions for the matrix elements of the relative kinetic operator, the Bloch operator and the non-local potential. The basis is defined as a set of $N$ functions $f_n(x)$ 
(see \cite{Hesse1998} and references therein), given by
\begin{equation}
f_n(x)=(-1)^{N+n}a^{-1/2}\sqrt{\frac{1-x_n}{x_n}}\frac{xP_N(2x/a-1)}{x-ax_n}\,,
\end{equation}
where $P_N(x)$ are Legendre polynomials, and $x_n$ satisfy
\begin{equation}
P_N(2x_n-1)=0\,.
\end{equation}
The Lagrange mesh associated with this basis consists of $N$ points $ax_n$ 
on the interval $[0,a]$ and satisfies the Lagrange condition 
\begin{equation}
f_{n'}(ax_n)=\frac{1}{\sqrt{a\lambda_n}}\delta_{nn'},
\label{LagCond}
\end{equation}
where the coefficients $\lambda_n$ are the weights corresponding to a
Gauss-Legendre quadrature approximation for the $[0,1]$ interval, i.e.
\begin{equation}
\int_0^1g(x)dx \sim \sum_{n=1}^N \lambda_ng(x_n)\,.
\end{equation}
Using the Lagrange conditions of Eq.~(\ref{LagCond}), it is straightforward to see that
within the Gauss approximation the Lagrange functions are orthogonal, i.e.
\begin{equation}
\int_0^a f_n(x) f_{n'}(x)dx\sim \delta_{nn'}.
\end{equation}

The matrix elements of the scattering matrix can then be used to calculate scattering phase shifts, cross sections and other reaction observables. As an example, the total cross section is given by
\begin{equation}
	\sigma_{\rm tot} = \frac{2\pi}{k^2_\nu}\sum_{J\pi\nu}\frac{2J+1}{(2I_1+1)(2I_2+1)}\left[1-{\rm Re}\!\left(S^{J\pi}_{\nu\nu}\right)\right]\,.
\end{equation}
  
Finally, while in the present discussion the $R$-matrix is computed on the basis of a microscopic theory of nuclei, this matrix can also be written in terms of a set of parameters which can be constrained phenomenologically by using the theory to fit experimental cross sections~\cite{R-matrix}. This version of the theory, known as phenomenological $R$-matrix, is often one of the first methods used to interpolate and extrapolate to low-energy cross sections important for astrophysics and other applications (see, e.g., Ref.~\cite{RevModPhys.89.035007}). For a review of $R$-matrix theory we refer the interested reader to Ref.~\cite{R-matrix}.

\section{Nuclear Reactions in Astrophysics}\label{NRA}

Nuclear reactions play an important role in astrophysics and cosmology. Lightest elements are synthesised in early universe in the Big Bang reaction chain. Heavier nuclei up to iron are produced by fusion, transfer, and radiative capture processes in the stars, while still heavier nuclei are produced by the slow, intermediate, or rapid neutron capture (i.e., s-process, i-process, r-process) in neutron rich environments, and by the proton capture in the rp-process~\cite{BertulaniIoP,BoydUCP,ThompsonCUP}.

Thermonuclear reaction rates are crucial for nuclear astrophysics. The number of nuclear reactions per volume and per second can be expressed as $r_{12}{=}\langle \sigma v \rangle n_1 n_2$ with target and projectile number densities $n_1, n_2$, respectively, $v$ the relative velocity of the two nuclei and $\sigma$ the cross section. When an equilibrium in the system is reached and the plasma environment can be characterized by a temperature $T$, the Maxwell-Boltzmann velocity distribution of interacting nuclei is assumed. The quantity $\langle \sigma v \rangle$ is then obtained as
\begin{equation}
    \langle \sigma v \rangle = \left(\frac{8}{\mu\pi}\right)^{1/2} (kT)^{-3/2} \int_0^\infty E \sigma(E) \; {\rm exp}(-E/kT) \;dE \; ,
\end{equation}
with $k$ the Boltzmann constant and $\mu$ the reduced mass of the target-projectile system. The nuclear cross section for charged particles is strongly suppressed at low energies due to Coulomb barrier. It is convenient to introduce the astrophysical $S$ factor
\begin{equation}\label{Sfact}
    S(E)=\sigma(E)\; E \; {\rm exp}(2\pi\eta) \; ,
\end{equation}
with $\eta{=}Z_1 Z_2 e^2/\hslash v$ the Sommerfeld parameter describing the $S$-wave barrier penetration. While the cross section decreases dramatically with the decrease of energy, the astrophysical S factor varies only moderately. Using Eq.~(\ref{Sfact}), the reaction rate $\langle \sigma v \rangle$ can be expressed as
\begin{equation}\label{reacrate}
    \langle \sigma v \rangle = \left(\frac{8}{\mu\pi}\right)^{1/2} (kT)^{-3/2} \int_0^\infty  S(E) \; {\rm exp}(-E/kT-d/E^{1/2}) \;dE \; ,
\end{equation}
where $d{=}(2\mu)^{1/2}\pi Z_1 Z_2 e^2/\hslash$. The integrant is suppressed at low energies due to the Coulomb repulsive interaction and at high energy due to the Boltzmann factor. The rather narrow energy range that contributes to the integral is called Gamow peak or Gamow window. The reaction rate will be strongly impacted by any resonance in the Gamow window. 

\subsection{Big Bang Nucleosynthesis}\label{bbn}

\begin{figure}
    \centering
    \includegraphics{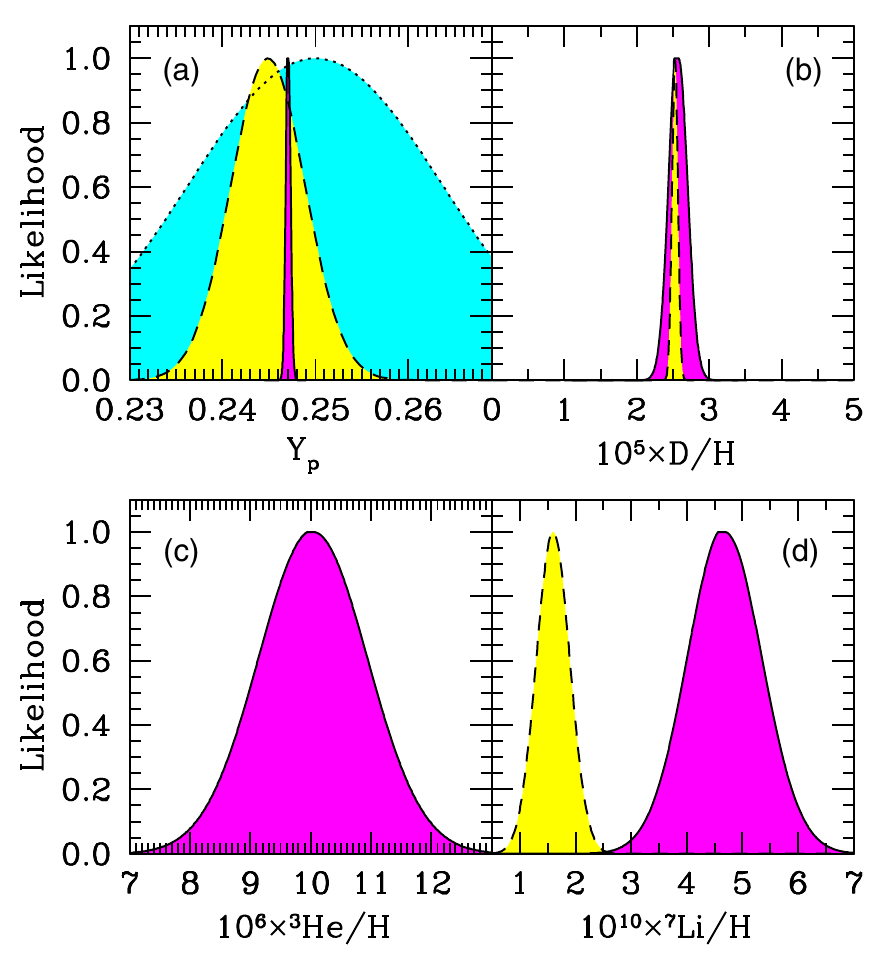}
    \caption{BBN light-element predictions for $^4$He (a), deuterium (b), $^3$He (c), $^7$Li (d),  using the CMB determination of the cosmic baryon density from the Planck data. The solid-lined, dark-shaded (purple) curves are the BBN predictions. The dashed-lined, light-shaded (yellow) curves show astronomical measurements of the primordial abundances, for all but $^3$He where reliable primordial abundance measures do not exist. For $^4$He, the dotted-lined, medium-shaded (cyan) curve shows the CMB determination of $^4$He. Reproduced from Ref.~\cite{RevModPhys.88.015004} where further details are given.}
    \label{fig:BBN}
\end{figure}

Primordial nucleosynthesis is believed to have taken place in the interval from roughly 10 seconds to 20 minutes after the Big Bang. The Big Bang Nucleosynthesis (BBN) theory predicts that roughly 25\% of the mass of the Universe consists of helium. It also predicts about 0.001\% deuterium, and even smaller quantities of lithium~\cite{RevModPhys.88.015004}. The prediction depends critically on the density of baryons (i.e., neutrons and protons) at the time of nucleosynthesis. The most recent determination of the baryon-to-photon ratio has been obtained from the Cosmic Microwave Background (CMB) {\it Planck} data~\cite{Planck2015}. It also depends on reaction rates of multiple  nuclear reactions involving light nuclei. We list the most important ones in Table~\ref{tab:BBN}.
\begin{table}[]
    \centering
    \begin{tabular}{cccccc}
        n(p,$\gamma$)$^2$H  & $^2$H(p,$\gamma$)$^3$He   & $^2$H(d,$\gamma$)$^4$He  & $^2$H(d,n)$^3$He    & $^2$H(d,p)$^3$H & $^3$He(n,p)$^3$H   \\
        $^3$H(d,n)$^4$He    & $^3$He(d,p)$^4$He        & $^2$H($\alpha$,$\gamma$)$^6$Li & $^6$Li(p,$\gamma$)$^7$Be & $^6$Li(p,$\alpha$)$^3$He & $^6$Li(n,$\alpha$)$^3$H \\
        $^3$H($\alpha$,$\gamma$)$^7$Li & $^3$He($\alpha$,$\gamma$)$^7$Be & $^7$Li(p,$\alpha$)$^4$He & $^7$Li(p,$\gamma$)$^8$Be & $^7$Be(n,p)$^7$Li & \\   
    \end{tabular}
    \caption{Reactions relevant for Big Bang Nucleosynthesis with d$\equiv ^2$H, $\alpha \equiv ^4$He. $^8$Be promptly decays to two $^4$He nuclei.}
    \label{tab:BBN}
\end{table}
While the BBN predictions of $^4$He and deuterium abundances are consistent with observations, the predictions for $^7$Li is three times higher than observed, see Fig.~\ref{fig:BBN}. This fact is often referred to as $^7$Li problem or puzzle. Given the $^7$Li discrepancy, it is imperative to understand well the underlying nuclear physics and in particular the reactions rates contributing to $^7$Li production. The BBN also predicts abundances of $^3$He and $^6$Li. However, observations of either of these isotopes is challenging and only limits are established. The BBN produces $^6$Li by the capture reaction $^2$H($\alpha$,$\gamma$)$^6$Li. There were recent claims of $^6$Li detection~\cite{Asplund_2006}, with isotope ratios as high as $^6$Li$/^7$Li${\lesssim}0.1$, which is orders of magnitude higher than the BBN prediction. This would constitute a ``$^6$Li problem". However, observation interpretations of Ref.~\cite{Asplund_2006} have been disputed~\cite{Cayrel2007,Lind2013}.

\subsection{Stellar Evolution}\label{stelevol}

\begin{figure}
    \centering
    \includegraphics[width=1.0\textwidth]{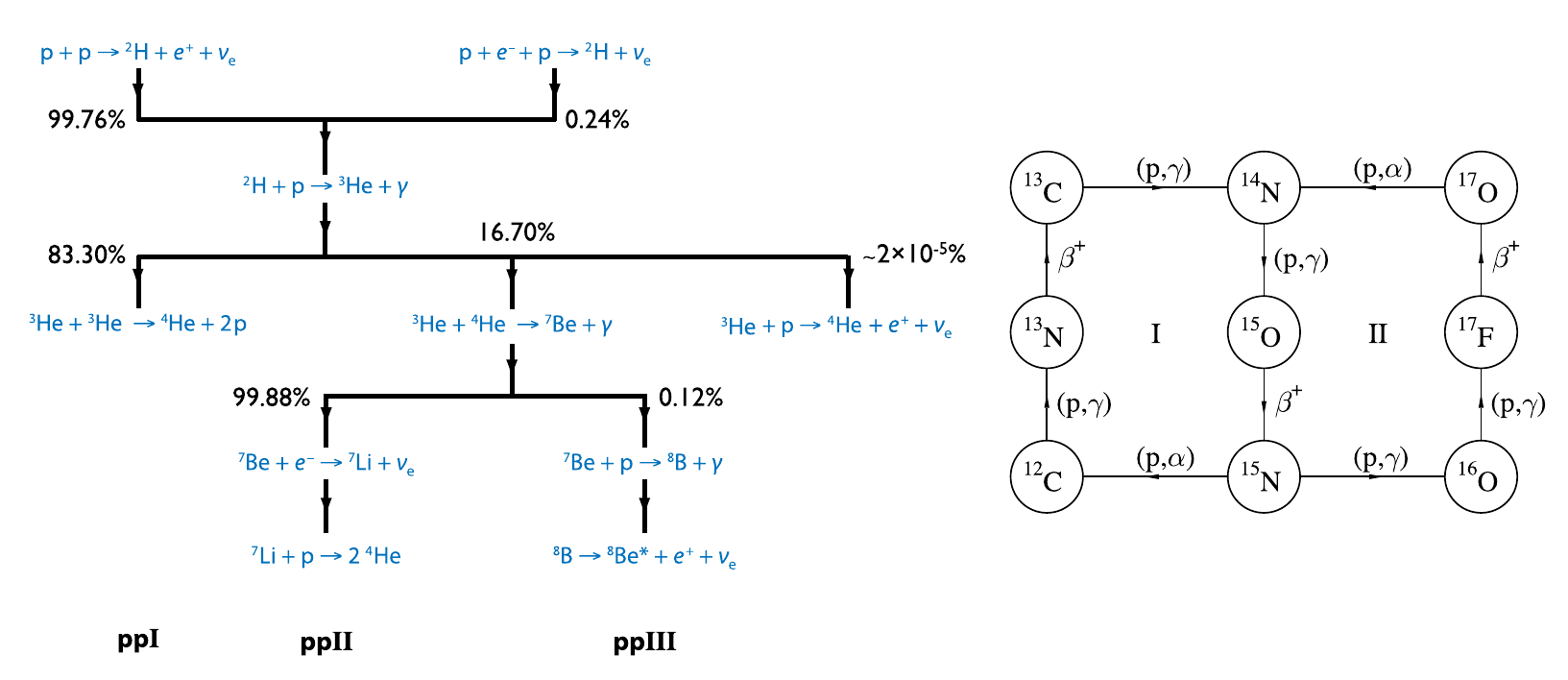}
    \caption{Left panel shows the cycles of the pp chain with branching percentages indicated. The right panel shows the CNO bicycle. Reproduced from Ref.~\cite{RevModPhys.83.195} where further details are given.}
    \label{fig:pp_CNO}
\end{figure}

Stars spend most of their lives burning the hydrogen in their cores. The hydrodynamic equilibrium is established with the gravitational force balancing the pressure created by the plasma thermal energy produced in the core of the star by nuclear reactions. There are two sets of hydrogen burning reactions, the proton-proton (pp) chain and the CNO cycles. Both sets of reactions transform four protons into an $\alpha$ particle ($^4$He nucleus). Corresponding individual nuclear reactions with pp chain branching percentages are presented in Fig.~\ref{fig:pp_CNO}. The slowest nuclear reaction in the pp chain is the proton-proton fusion p+p$\rightarrow$ d+e$^+$+$\nu_e$, which is due to the weak interaction. As the cross section of this reaction for low-energy protons is very small, of the order of $10^{-23}$~b, the average lifetime of protons in the Sun due to the transformation to deuterons by this reaction is about $10^{10}$ years. The slowest CNO reaction is the electromagnetic proton capture on $^{14}$N, $^{14}$N(p,$\gamma$)$^{15}$O. In the CNO cycles, the carbon, oxygen, and nitrogen nuclei are catalyzers for the nuclear processes. Due to the larger Coulomb repulsion of the catalyzing nuclei, the CNO cycles occur at higher temperatures. In the Sun, the pp chain dominates but in stars with higher temperatures, the CNO cycles are more important. 

The hydrogen burning reactions generate neutrinos of various energies. The $^7$Be(p,$\gamma$)$^8$B capture reaction is particularly significant as the $^8$B that it produces is responsible for nearly all of the yield in the solar neutrino detectors. The electron neutrino from the $^8$B beta decay has energy of the order of 15 MeV, much higher than neutrinos emitted in other hydrogen burning reaction (except the low-intensity $^3$He+p weak capture, see Fig.~\ref{fig:pp_CNO}). Consequently, it plays a significant role in experimental verification of the standard solar model~\cite{Bahcall_2005,SERENELLI201341} and in the discovery of the neutrino oscillations~\cite{PhysRevLett.87.071301}.

If temperature in the core of a star increases, helium burning can occur. Two sets of reactions dominate helium burning. The first converts the $^4$He into $^{12}$C in the triple-$\alpha$ reaction that proceeds mostly sequentially: $\alpha{+}\alpha\rightarrow ^8$Be forming the $^8$Be g.s. resonance with a lifetime of the order of $10^{-16}$~s followed by the radiative capture of the third $\alpha$ particle, $^8$Be($\alpha$,$\gamma$)$^{12}$C, which is strongly enhanced by the $0^+$ resonance, the so-called Hoyle state~\cite{Hoyle1954}, just above the triple-$\alpha$ threshold in $^{12}$C. Once $^{12}$C builds up, $^{16}$O can be produced by the $\alpha$ capture reaction $^{12}$C($\alpha$,$\gamma$)$^{16}$O. Reaction rates of these two reactions ultimately determine the carbon-to-oxygen ratio which subsequently impacts production of heavier elements. 

With a further temperature increase in the core of a star, burning of heavier elements such as $^{12}$C, $^{16}$O, $^{20}$Ne, $^{28}$Si will occur and produce elements up to iron.

Accurate knowledge of nuclear cross sections of reactions involved in various stellar burning cycles in particular in the energy range relevant for the Gamow window of the respective reactions is crucial for the validation of the standard solar model, for understanding the fluxes of solar neutrinos, as well as for our understanding of abundances of elements in the universe.   

\section{\textit{Ab Initio} Calculations of Reactions Important for Astrophysics}

A predictive nuclear theory would greatly help our understanding of nuclear reactions important for astrophysics. Typically, capture, transfer or other reactions take place in the Cosmos at energies much lower than those accessible by experiments. A well-known example is provided by the triple-$\alpha$ and $^{12}$C($\alpha,\gamma$)$^{16}$O radiative capture reactions discussed in the previous section. The ratio of the thermonuclear reaction yields for these two processes determines the carbon-to-oxygen ratio at the end of helium burning with important consequences for the production of all species made in subsequent burning stages in the stars. At stellar energies ($\approx300$ keV) radiative capture rates are too small to be measured in the laboratory. Thus, measurements are performed at higher energies (see, e.g., the experiment reported in Ref.~\cite{PhysRevLett.97.242503}) and extrapolations to the low energy of interest using theory are unavoidable. Theoretical extrapolations are, however, challenging due to the influence of several resonances. A fundamental theory would be of great use here.

The NCSMC approach still needs to be further developed to address the complex nature of the triple-$\alpha$ and the $^{12}$C($\alpha,\gamma$)$^{16}$O reactions. However, the first scattering calculations involving $\alpha$-particle projectile have been performed~\cite{Kravvaris:2020cvn}. Still, there have been numerous applications of the NCSMC and the NCSM/RGM to reactions relevant for astrophysics. In this section, we discuss recent calculations of the deuterium (D) tritium (T) and $^3$He(d,p)$^4$He fusion, the $^3$He($\alpha$,$\gamma$)$^7$Be, $^7$Be(p,$\gamma$)$^8$B, and $^8$Li(n,$\gamma$)$^9$Li radiative capture reactions.

While the cross section of the fusion or rather transfer reactions is directly obtained by the application of the formalism presented in Subsection Unified Description of Bound and Continuum States: the No-Core Shell Model with Continuum, the cross section of the radiative capture reactions requires calculations of matrix elements of electromagnetic operators with an initial scattering state, e.g., $^7$Be+p and a final bound state, e.g., $^8$B. The capture proceeds dominantly by the electric dipole (E1) radiation. The corresponding transition operator can be cast in the form
\begin{equation}\label{eq:E1}
\hat{D}= e \sum_{i=1}^A \frac{1+\tau_i^z}{2} (\vec{r}_i-\vec{R}^{(A)}_{\rm c.m.}) \;,
\end{equation}  
with $\tau_i^z$ and $\vec{r}_i-\vec{R}^{(A)}_{\rm c.m.}$ representing the isospin third component ($\tau_i^z{=}1(-1)$ for proton(neutron)) and center of mass frame coordinate of the {\it i}th nucleon. This form of the E1 transition operator includes the leading effects of the meson-exchange currents through the Siegert's theorem. The magnetic dipole (M1) and electric quadrupole (E2) radiation may also play a role in particular at and around resonances. In the NCSMC, wave functions have two components, a square-integrable NCSM part and a cluster NCSM/RGM part, see Eq.~(\ref{NCSMC_wav}). Consequently, the operator matrix element has four contributions and its evaluation is rather involved. While the contribution with the NCSM eigenstates, Eqs.~(\ref{NCSM_wav}) and (\ref{NCSMC_wav}), in the initial and the final state is straightforward to obtain, the other contributions with cluster states in the initial and/or the final state require more attention. To evaluate those terms, we rewrite the E1 operator (\ref{eq:E1}) as an operator acting exclusively on the first $A{-}a$ nucleons (pertaining to the first cluster or target), an operator acting exclusively on the last $a$ nucleons (belonging to the second cluster or projectile), and, finally, an operator acting on the relative motion wave function between target and projectile:
\begin{eqnarray}\label{eq:E1cl}
\hat{D}&=& e \sum_{i=1}^{A-a} \frac{1+\tau_i^z}{2} (\vec{r}_i-\vec{R}^{(A-a)}_{\rm c.m.}) + e \sum_{i=A-a+1}^{A} \frac{1+\tau_i^z}{2} (\vec{r}_i-\vec{R}^{(a)}_{\rm c.m.}) \nonumber \\
&+& e \frac{Z_{(A-a)}a-Z_{(a)}(A-a)}{A}\vec{r}_{A-a,a}\;.
\end{eqnarray}
Here, $\vec{R}^{(A-a)}_{\rm c.m.}$ and $\vec{R}^{(a)}_{\rm c.m.}$ are the centers of mass of the $(A-a)$- and $a$-nucleon systems, respectively, $\vec{r}_{A-a,a}=\vec{R}^{(A-a)}_{\rm c.m.}{-}\vec{R}^{(a)}_{\rm c.m.}$, while $Z_{(A-a)}$ and $Z_{(a)}$ represent respectively the charge numbers of the target and of the projectile. The dominant contribution is due to the relative motion term that we are able to evaluate without any approximations. It involves integration of wave functions of the relative motion that needs to be carried out to large distances. For example, in the case of $^7$Be(p,$\gamma$)$^8$B capture the cluster-cluster term needs to be integrated up to 250 fm as the $^8$B g.s. is very weakly bound. Matrix elements of the operators acting on the target and projectile we evaluate approximately using completeness of the target and projectile or of the composite states. The E2 and the orbital part of the M1 operator can be dealt with in an analogous way. As to the spin part of the M1 operator($g_p \sum_{i=1}^A \frac{1+\tau_i^z}{2}\vec{s}_i+g_n \sum_{i=1}^A \frac{1-\tau_i^z}{2}\vec{s}_i$), we have developed a capability to calculate its matrix elements without any approximations using a technique similar to that applied to calculate integration kernels of the NN interaction. The same way we are able to compute matrix elements of the $\beta$ decay Gamow-Teller ($g_A \sum_{i=1}^A \vec{\sigma}_i \tau_i^+$) and Fermi operators ($\sum_{i=1}^A\tau_i^+$).  

\subsection{Deuterium (D) Tritium (T) and D$^3$He Fusion}\label{sec:DT}

The $^3$H(d,n)$^4$He and $^3$He(d,p)$^4$He reactions are leading processes in the primordial formation of the very light elements (mass number, $A\le7$), affecting the predictions of Big Bang Nuleosynthesis for light nucleus abundances~\cite{1475-7516-2004-12-010}. With its low activation energy and high yield, $^3$H(d,n)$^4$He is also the easiest reaction to achieve on Earth, and is pursued by research
facilities directed toward developing fusion power by either magnetic ({\em e.g.}\ ITER) or inertial ({\em e.g.}\ NIF) confinement.   
The cross section for the DT fusion is well known experimentally, while more uncertain is the situation for the branch of this reaction,
$^3$H(d,$\gamma$ n)$^4$He that produces $17.9$ MeV $\gamma$-rays~\cite{PhysRevLett.53.767,PhysRevC.47.29} and that is being considered as a possible plasma diagnostics in modern fusion experiments. Larger uncertainties dominate also
the $^3$He(d,p)$^4$He reaction that is known for presenting considerable electron-screening effects
at energies accessible by beam-target experiments. Here, the electrons bound to the target, usually a neutral atom or molecule,
lead to increasing values for the reaction-rate with decreasing energy, effectively preventing direct access to the astrophysically relevant bare-nucleus cross section. Consensus on the physics mechanism behind this enhancement is not been reached yet~\cite{Kimura2005229}, largely because of the difficulty of determining the absolute value of the bare cross section.
 
Past theoretical investigations of these fusion reactions include various $R$-matrix analyses of experimental data at higher energies~\cite{PhysRevLett.59.763,PhysRevC.56.2646,PhysRevC.75.027601,Descouvemont2004203}  as well as microscopic calculations with phenomenological interactions~\cite{PhysRevC.41.1191,Langanke1991,PhysRevC.55.536}. However, in view of remaining experimental challenges and the large role played by theory in extracting the astrophysically important information, it is highly desirable to achieve a microscopic description of the $^3$H(d,n)$^4$He and $^3$He(d,p)$^4$He fusion reactions that encompasses the dynamic of all five nucleons and is based on the fundamental underlying physics: the realistic interactions among nucleons and the structure of the fusing nuclei.

\begin{figure}
    \centering
    \begin{subfigure}[b]{0.49\textwidth}
         \centering
         \includegraphics[width=\textwidth]{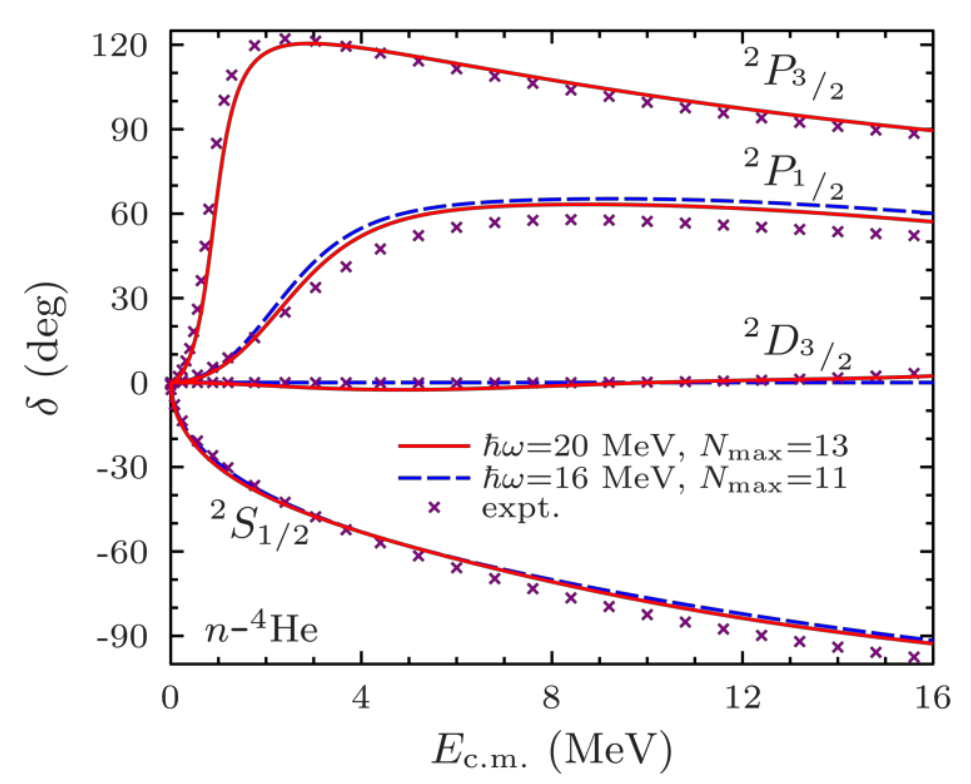}
%         \caption{$y=x$}
%         \label{fig:y equals x}
     \end{subfigure}
            \hfill
    \begin{subfigure}[b]{0.49\textwidth}
         \centering
         \includegraphics[width=\textwidth]{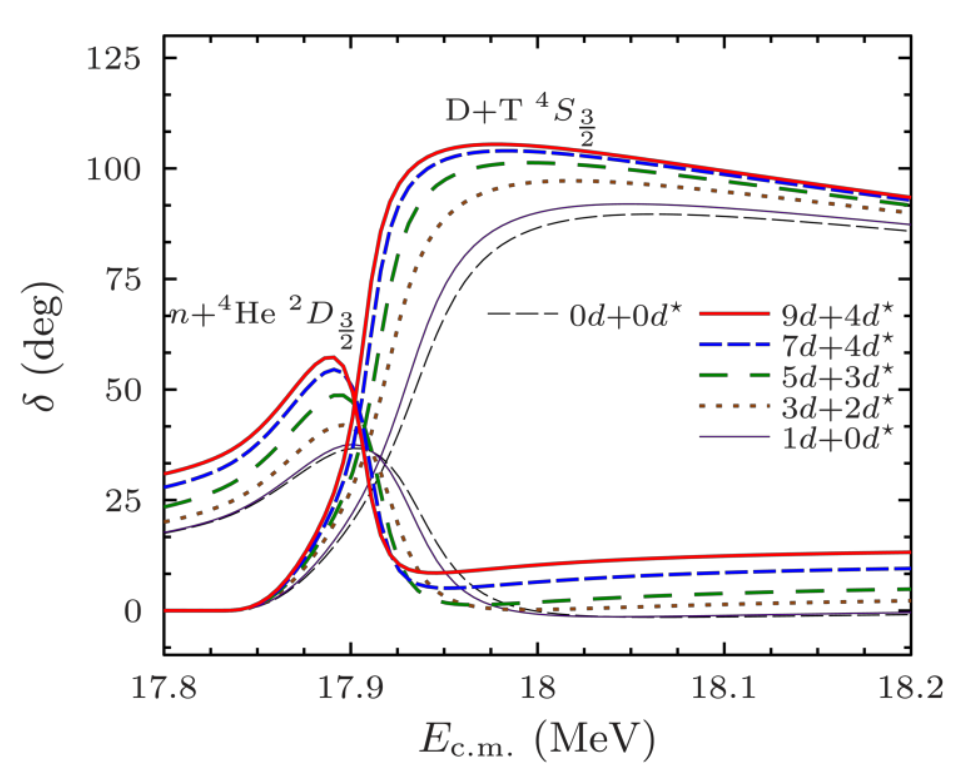}
%         \caption{$y=x$}
%         \label{fig:y equals x}
     \end{subfigure}
    \caption{Left panel: Neutron-$^4$He phase shifts below the fusion reaction threshold. Results in the $^2S_{1/2}$,$^2P_{1/2}$,$^2P_{3/2}$, and $^2D_{3/2}$ partial waves obtained in the NCSMC using two HO frequencies and basis sizes compared to an accurate R-matrix parametrization of experimental data (G. M. Hale, personal communication). Right panel: Diagonal phase shifts in the entrance and exit channels of the DT fusion. Convergence of the (real part of the) D+T $^4S_{3/2}$ and n+$^4$He $^2D_{3/2}$ diagonal phase shifts (characterizing, respectively, the entrance and exit scattering states of the DT fusion reaction) in the $J^\pi{=}3/2^+$ channel with increasing number of positive-energy eigenstates of the deuterium, as obtained within the NCSMC approach. Adopted from Ref.~\cite{Hupin2019} where further details are given.}
    \label{fig:He5_phaseshift}
\end{figure}

\begin{figure}
    \centering
    \begin{subfigure}[b]{0.49\textwidth}
         \centering
         \includegraphics[width=\textwidth]{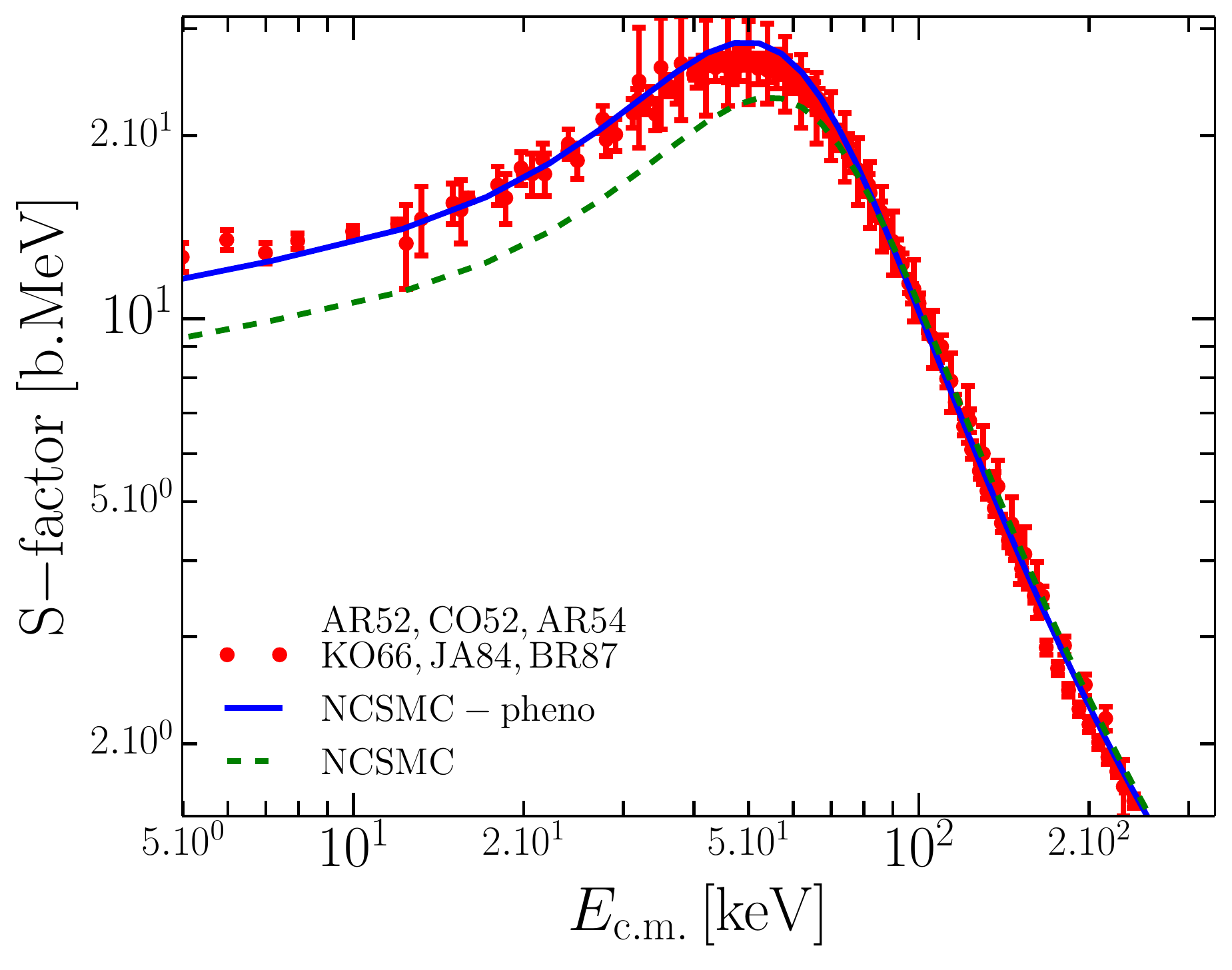}
     \end{subfigure}
            \hfill
    \begin{subfigure}[b]{0.49\textwidth}
         \centering
         \includegraphics[width=\textwidth]{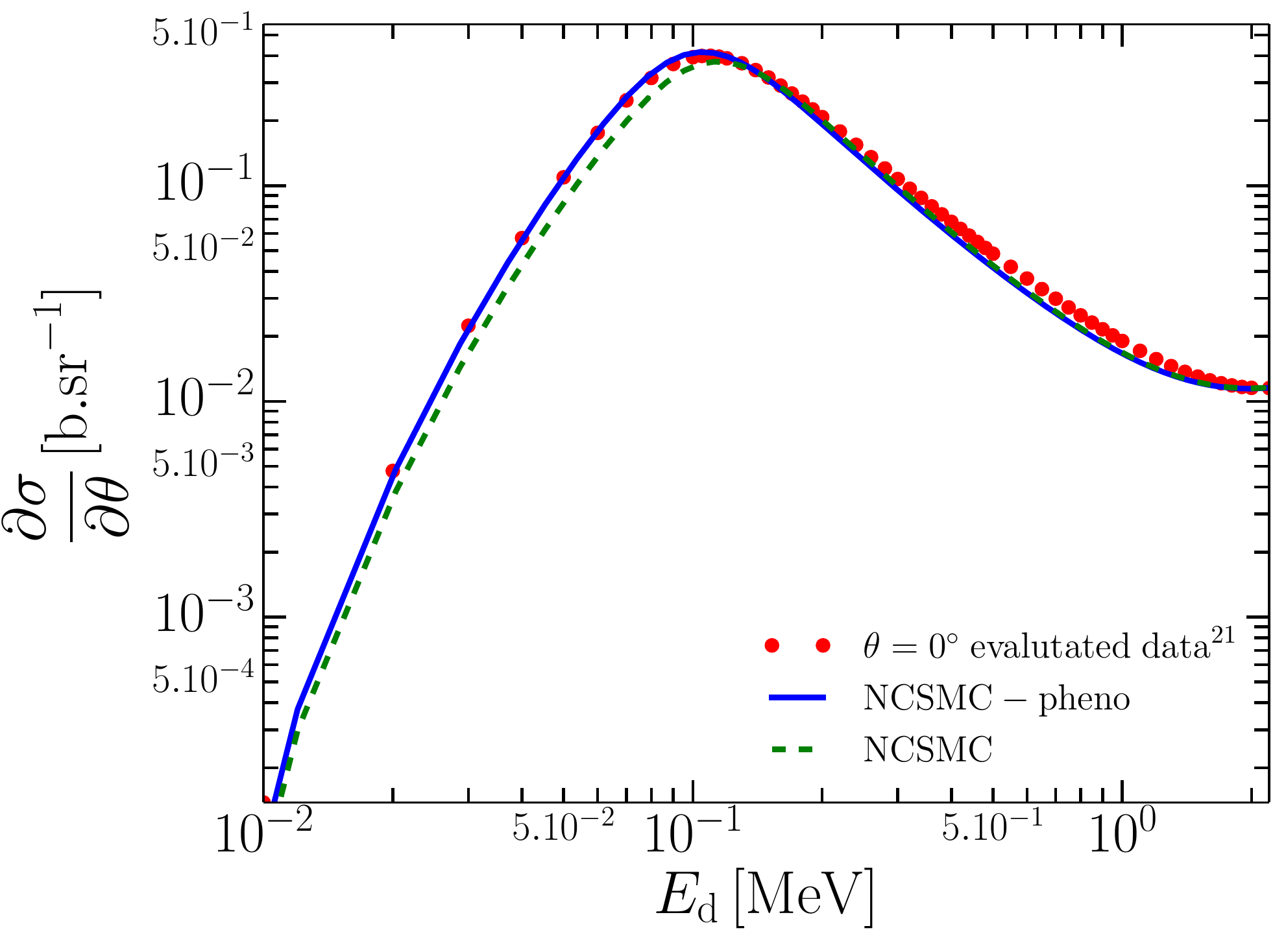}
     \end{subfigure}
         \begin{subfigure}[b]{0.49\textwidth}
         \centering
         \includegraphics[width=\textwidth]{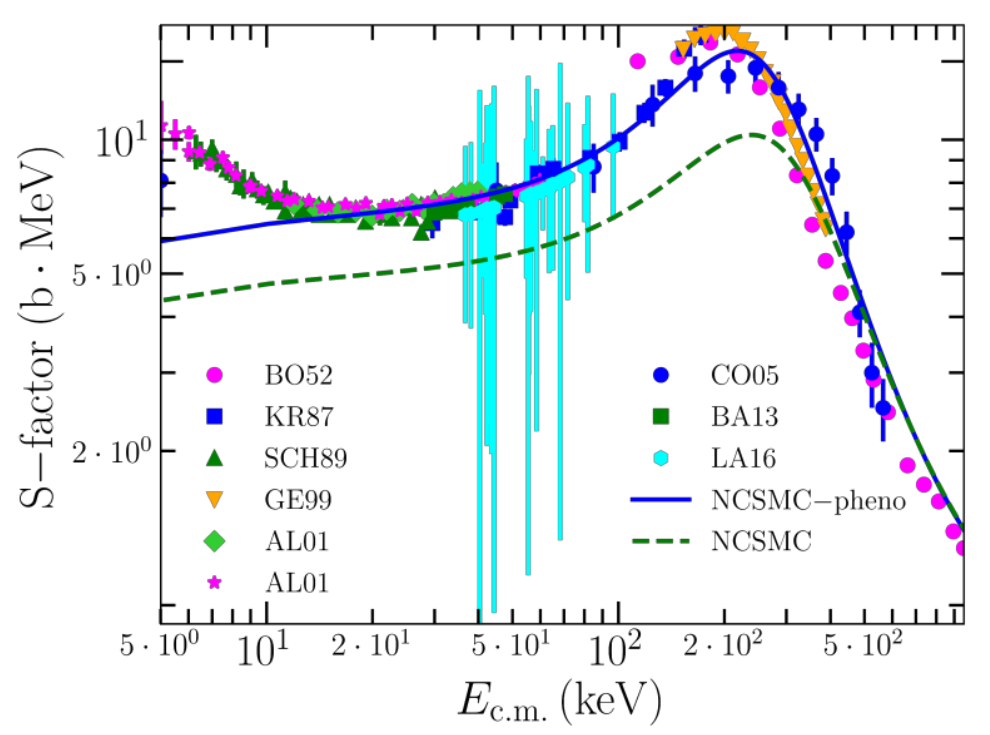}
     \end{subfigure}
            \hfill
    \begin{subfigure}[b]{0.49\textwidth}
         \centering
         \includegraphics[width=\textwidth]{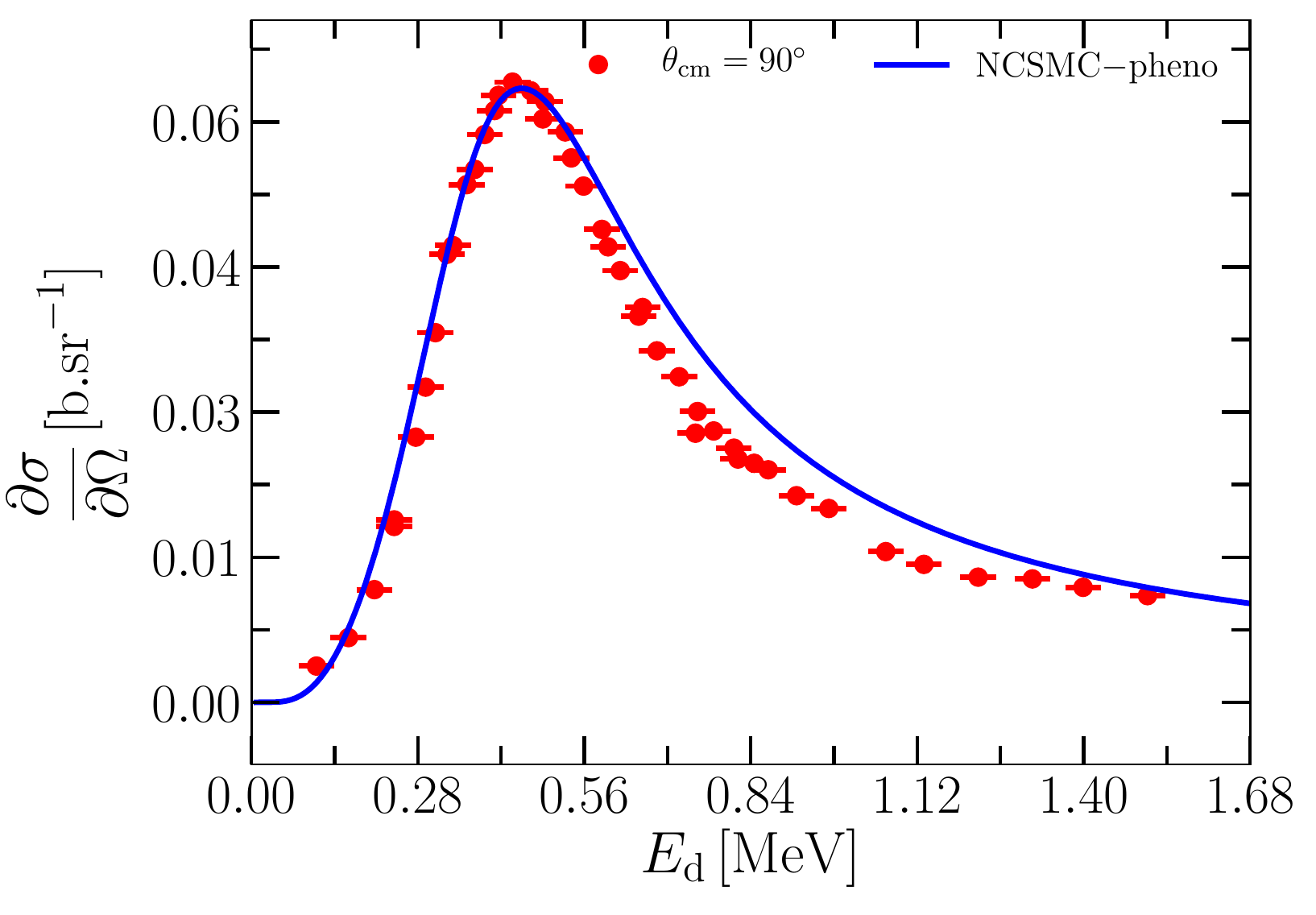}
     \end{subfigure}
    \caption{Top (bottom) left panel: Astrophysical S factor of the $^3$H(d,n)$^4$He ($^3$He(d,p)$^4$He) reaction as a function of the energy in the c.m. frame, $E_{\rm c.m.}$, compared to available experimental data. Top (bottom) right panel: Angular differential cross section as a function of the deuterium incident energy, $E_{\rm d}$, for the $^3$H(d,n)$^4$He ($^3$He(d,p)$^4$He) at the c.m. scattering angle of $\theta_{\rm c.m.}{=}0^o$ ($90^o$) compared to the evaluated data. The “NCSMC” and “NCSMC-pheno” stand for the results of the present calculations before and after a phenomenological correction to the position of the $3/2^+$ resonance. Adopted from Ref.~\cite{Hupin2019} where further details are given.}
    \label{fig:A5_xsect}
\end{figure}

We made the first step in this direction by performing NCSM/RGM calculations using a realistic NN interaction~\cite{Navratil2012}. A much more advanced NCSMC investigation of the DT fusion was presented in Ref.~\cite{Hupin2019}. These calculations include both the $^4$He+n and the $^3$H+d (or $^3$He+d) mass partitions in the cluster part of the NCSMC trial wave function given in Eqs.(\ref{eq:NCSMC-eq}) and (\ref{eq:rgm-basis}). While the main focus was on the calculation of observables for the polarized D and T nuclei that have not been measured yet, phase shifts, cross sections as well as results for the mirror $^3$He(d,p)$^4$He system were presented. Chiral NN interaction from Ref.~\cite{Entem2003} and chiral 3N interaction from Ref.~\cite{Navratil2007} served as input for these calculations. NCSMC calculated neutron-$^4$He phase shifts are compared to an R-matrix fit to experimental data are shown in the left panel of Fig.~\ref{fig:He5_phaseshift}. Note that we use the spectroscopic notation $^{2s+1}l_J$ for the partial waves with $s$ the channel spin, $l$ the orbital momentum with $S$ for $l{=}0$, $P$ for $l{=}1$, $D$ for $l{=}2$, etc., and $J$ the total angular momentum with the couplings defined in Eq.~(\ref{eq:rgm-basis}). To asses the stability and convergence of the NCSMC calculation, the HO frequency and the basis size of the NCSM eigenstate expansion were varied. It is clear that the NCSMC with the selected chiral NN+3N Hamiltonian is capable to reproduce experiment quite well. At low energy, we observe two $P$-wave resonances corresponding to the $^5$He $3/2^-$ ground state and a broad $1/2^-$ excited state. At the energy of about 17.6 MeV (17.9 MeV in the calculation), we find a narrow $3/2^+$ resonance manifested in the $^2D_{3/2}$ n+$^4$He partial wave as well as in the d+$^3$H $^4S_{3/2}$ partial wave as shown in the right panel of Fig.~\ref{fig:He5_phaseshift}. A convergence with respect to the number of D pseudo-excited states that approximate the p+n+$^3$H continuum in the NCSMC calculations is shown. This $3/2^+$ resonance is responsible for a strong enhancement of the DT fusion cross section shown in the top two panels of Fig.~\ref{fig:A5_xsect}. The DT fusion or rather a transfer reaction proceeding through this resonance resuls in a release of 17.6 MeV of energy distributed to the emitted 14.1 MeV fast neutron and a 3.5 MeV $\alpha$ particle. 

In Fig.~\ref{fig:A5_xsect}, we present the astrophysical S factor related to the cross section by Eq.~(\ref{Sfact}) and the angular differential cross section of the $^3$H(d,n)$^4$He reaction and its mirror $^3$He(d,p)$^4$He. The effect of the $3/2^+$ resonance is clearly manifested. Overall, we find a very good agreement with the experimental data once we adjust the calculated position of the $3/2^+$ resonance centroid. This adjustment is quite small, only -5 keV for the DT case. However, its impact on the cross section is quite visible. The adjusted calculation dubbed NCSMC-pheno is performed by varying the input NCSM eigenstate energies, in the present case a single eigenvalue, entering the ${\cal E}$ term of Eq.~(\ref{eq:NCSMC-eq}). The experimental astrophysical S factor of the $^3$He(d,p)$^4$He reaction shows an enhancement at low energies as seen in the bottom left panel of Fig.~\ref{fig:A5_xsect}. This is due to the electron screening of the target nuclei in the beam-target experiments. Such a screening is not present in the stellar or primordial plasma. Consequently, it is important to know the astrophysical S factor of bare nuclei (without electrons) such as that obtained in the NCSMC calculation.   

\begin{figure}
    \centering
    \includegraphics[width=0.8\textwidth]{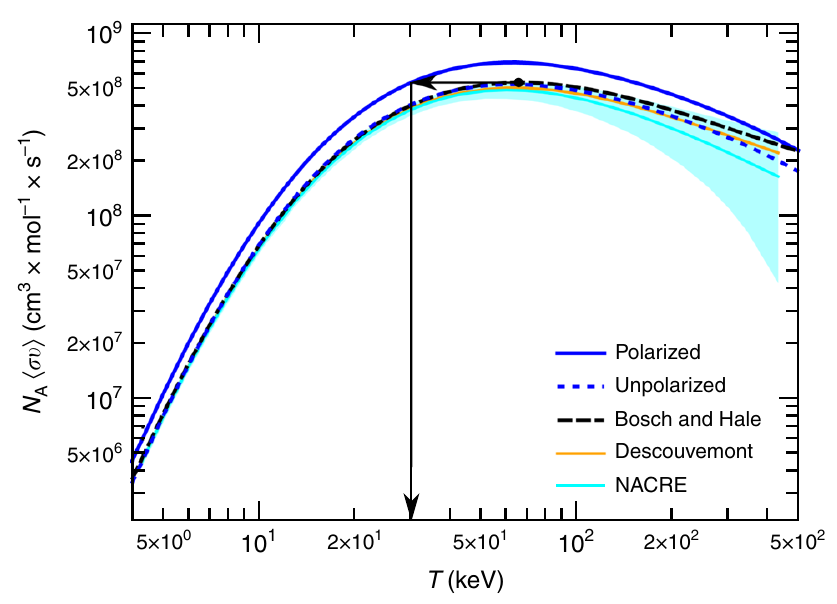}
    \caption{DT reaction rate with and without polarization. Comparison between the NCSMC computed DT reaction rate ($N_A \langle\sigma v\rangle$, with $N_A$ the Avogadro number) for unpolarized and polarized fuel with aligned spins as a function of the temperature, T. A phenomenological correction to the position of the $3/2^+$ resonance (dubbed NCSMC-pheno) has been applied. Also shown for comparison are the unpolarized reaction rates obtained from the widely adopted parametrization of the DT fusion cross section of Bosch and Hale, from the R-matrix fit of Descouvemont and from the NACRE compilation. The arrows in the figure show that, with polarization, a reaction rate of equivalent magnitude as the apex of the unpolarized reaction rate is reached at lower temperatures. Adopted from Ref.~\cite{Hupin2019} where further details are given.}
    \label{fig:DT_rate}
\end{figure}

Fifty years ago, it was estimated~\cite{PhysRevLett.49.1248} that, in the ideal scenario in which the spins of the D (g.s. spin and parity $1^+$) and T (g.s. spin and parity $1/2^+$) reactants are perfectly aligned in a total-spin $3/2^+$ configuration and assuming that the reaction is isotropic, one could achieve an enhancement of the cross section by a factor of 1.5, thus improving the economics of fusion energy generation. Due to technical complexity, no experimental measurement of the polarized cross section exits. The {\it ab initio} NCSMC calculation of Ref~\cite{Hupin2019} provides theoretical confirmation of this enhancement. In Fig.~\ref{fig:DT_rate}, we show the NCSMC calculated reaction rate with polarized and unpolarized DT nuclei. The unpolarized rate agrees well with rates obtained by evaluations of experimental data. The polarized reaction rate shows the same shape, albeit globally enhanced by a factor of ~1.32, in agreement with the approximate estimate for the chosen polarization of 80\%. This result follows from the rather slow variation of the enhancement factor of the reaction cross section as a function of the energy in the narrow Gamow window (deuteron incident energies below a few hundred keV). It is interesting to note that with polarization a reaction rate of equivalent magnitude as the apex of the unpolarized reaction rate is reached at lower temperatures, that is less than 30 keV compared to 65 keV (where both rates peak), as highlighted in Fig.~\ref{fig:DT_rate} by the arrows. As a naive illustration, this means that by using polarized DT fuel the output of a standard fusion reactor could either be enhanced by 32\% or its operational temperature decreased by as much as 45\%.

\subsection{$^3$He($\alpha,\gamma$)$^7$Be and $^3$H($\alpha,\gamma$)$^7$Li Radiative Capture Reactions}

The ${^3{\rm He}}(\alpha,\gamma){^7{\rm Be}}$ and ${^3{\rm H}}(\alpha,\gamma){^7{\rm Li}}$ radiative-capture processes hold great astrophysical significance. 
Their reaction rates for collision energies between  $\sim$20 and 500~keV  in the center-of-mass (c.m.)\ frame are essential to calculate the primordial ${^7{\rm Li}}$ abundance in the universe~\cite{PhysRevLett.82.4176,Nollett2001b}. 
In addition, standard solar model predictions for the fraction of pp-chain branches resulting in ${^7{\rm Be}}$ versus ${^8{\rm B}}$ neutrinos depend critically on the ${^3{\rm He}}(\alpha,\gamma){^7{\rm Be}}$ astrophysical $S$ factor at about 20~keV c.m.\ energy~\cite{RevModPhys.70.1265,RevModPhys.83.195}. Because of the Coulomb repulsion between the fusing nuclei, these capture cross sections are strongly suppressed at such low energies and thus hard to measure directly in a laboratory.

Concerning the ${^3{\rm He}}(\alpha,\gamma){^7{\rm Be}}$ radiative capture, experiments performed by several groups in the last two decades have led to quite accurate cross-section determinations for collision energies between about 90~keV and 3.1~MeV in the c.m.\ frame~\cite{Broggini2006,PhysRevC.75.065803,PhysRevLett.102.232502,BORDEANU20131}. 
However, theoretical models or extrapolations are still needed to provide the capture cross section at solar energies~\cite{Leva_2016}.
In contrast, experimental data are less precise and also much less extensive for the ${^3{\rm H}}(\alpha,\gamma){^7{\rm Li}}$ radiative capture. The most recent experiment was performed almost thirty years ago resulting in measurements at collision energies between about 50~keV and 1.2~MeV in the c.m.\ frame~\cite{PhysRevC.50.2205}.

Theoretically, these radiative captures have also generated much interest: from the development of pure external-capture models in the early 60's to the microscopic approaches from the late 80's up to now~\cite{RevModPhys.83.195,Nollett2001b,Neff2011}. 
However, no parameter-free approach is able to simultaneously reproduce the latest experimental ${^3{\rm He}}(\alpha,\gamma){^7{\rm Be}}$ and ${^3{\rm H}}(\alpha,\gamma){^7{\rm Li}}$ astrophysical $S$ factors.

\begin{figure}
    \centering
    \begin{subfigure}[b]{0.49\textwidth}
         \centering
         \includegraphics[width=\textwidth]{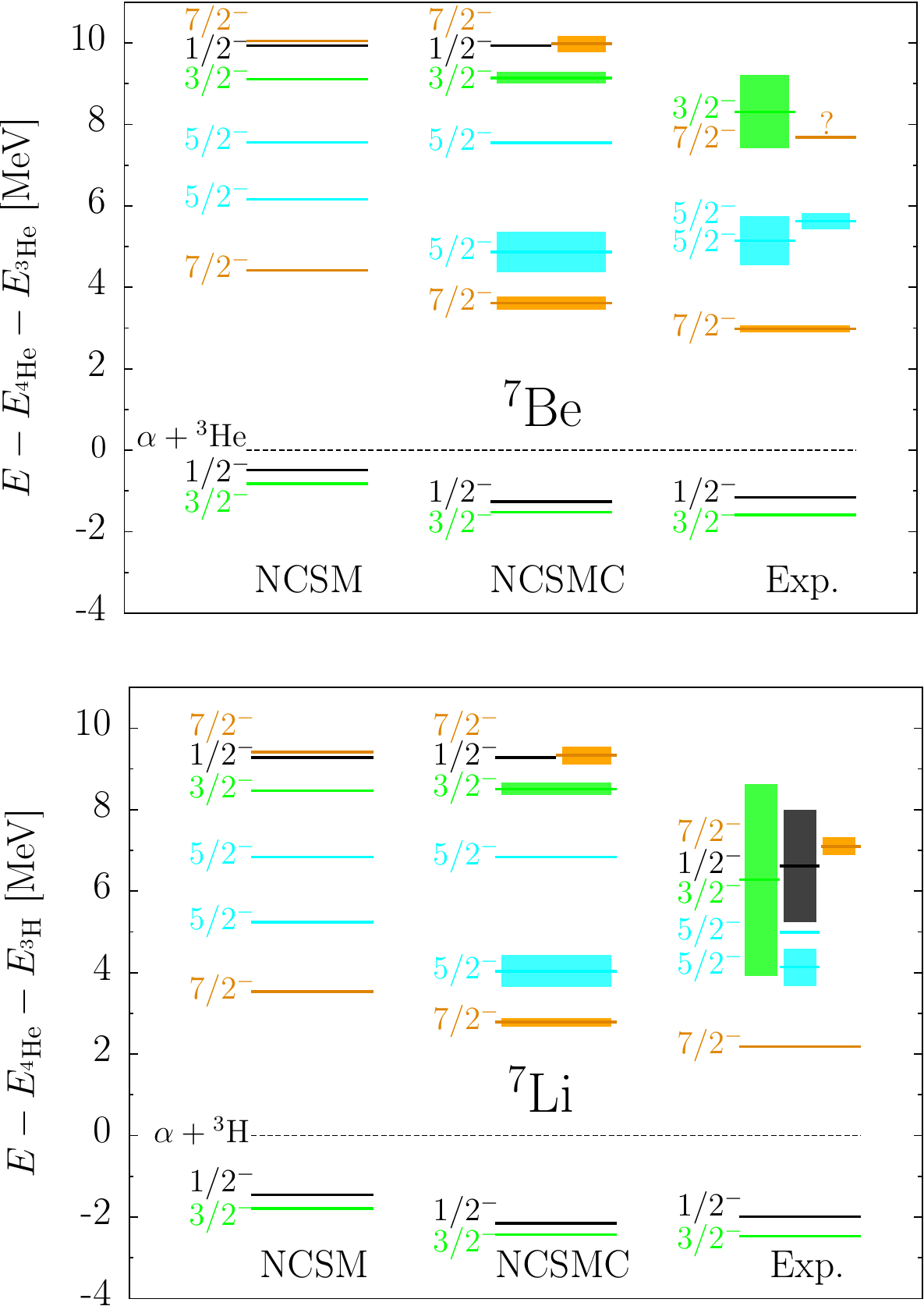}
     \end{subfigure}
            \hfill
    \begin{subfigure}[b]{0.49\textwidth}
         \centering
         \includegraphics[width=\textwidth]{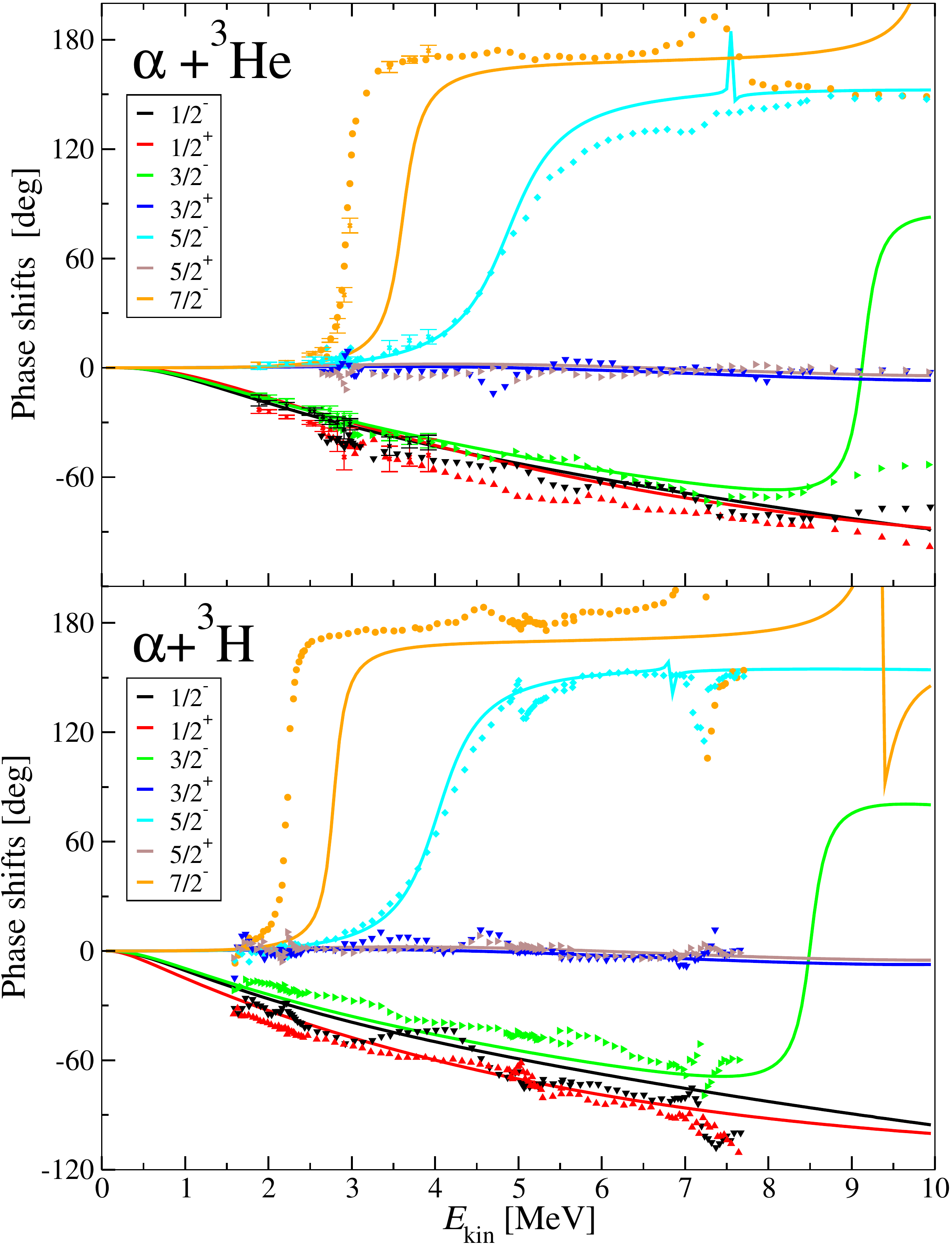}
     \end{subfigure}
    \caption{Left panels: The $^7$Be and $^7$Li spectra obtained from the NCSM and NCSMC approaches compared to experiment. Rectangles symbolize the widths of resonances. Right panels: The $\alpha{+}^3$He and $\alpha{+}^3$H elastic phase shifts obtained from the NCSMC approach and from experiments. Energies are given with respect to the $\alpha{+}^3$He/$^3$H threshold. Adopted from Ref.~\cite{DohetEraly2016} where further details are given.}
    \label{fig:7Be7Li_phaseshift}
\end{figure}

{\it Ab initio} NCSMC was applied recently to study $\alpha{+}^3$He and $\alpha{+}^3$H scattering, the structure of low-lying states of $^7$Be and $^7$Li, and to calculate the astrophysical S factor~\cite{DohetEraly2016}. As input, the chiral NN interaction of Ref.~\cite{Entem2003} renormalized by the SRG transformation with the flow parameter $\Lambda{=}2.15\; {\rm fm}^{-1}$ was used. As seen in the left panels of Fig.~\ref{fig:7Be7Li_phaseshift}, this interaction provides a very good description of the lowest states of $^7$Be and $^7$Li. It should be noted that the NCSM eigenstates shown in the leftmost columns serve as input to the NCSMC calculation presented in the middle columns. The NCSMC results clearly improve on the NCSM as to the positions of the bound states and the low-lying resonances and most importantly, they predict also widths of the resonances unlike the NCSM limited by its HO basis. The $\alpha{+}^3$He and $\alpha{+}^3$H phase shifts obtained from the NCSMC approach are presented together with experimental data in the right panels of Fig.~\ref{fig:7Be7Li_phaseshift}. The resonance energies and their widths were obtained by calculating the first derivative of the phase shifts: The resonance energy is obtained from the inflection point where the first derivative of the phase shift is maximal and the width is subsequently computed from the phase shifts according to
\begin{equation}\label{eq:reswidth}
  \Gamma=\left. \frac{2}{{\rm d} \delta(E_{\rm kin}) / {\rm d} E_{\rm kin}}\right|_{E_{\rm kin}=E_R}\,,
\end{equation}
where $E_R$ is the resonance centroid, evaluated as discussed above, and the phase shift is expressed in radians. $E_{\rm kin}$ is the kinetic energy in the c.m. We can see that the calculation overpredicts the position of the $7/2^-$ resonance and matches quite well the first $5/2^-$ resonance. For the sake of clarity, the jump of $+180^o$ in the phase shifts at the second $5/2^-$ and $7/2^-$ resonance energies are not displayed. The second $5/2^-$ resonance in $^7$Be ($^7$Li) is known to be dominated by the $^6$Li+p ($^6$Li+n) mass partition not included in the discussed calculations. Consequently, its position is missed as seen in Fig.~\ref{fig:7Be7Li_phaseshift}. It should be noted that new $\alpha$+$^3$He elastic scattering data from the TRIUMF SONIK experiment are now available~\cite{Paneru2020}. Those are, however, not included in Fig.~\ref{fig:7Be7Li_phaseshift}.

\begin{figure}
    \centering
    \begin{subfigure}[b]{0.49\textwidth}
         \centering
         \includegraphics[width=\textwidth]{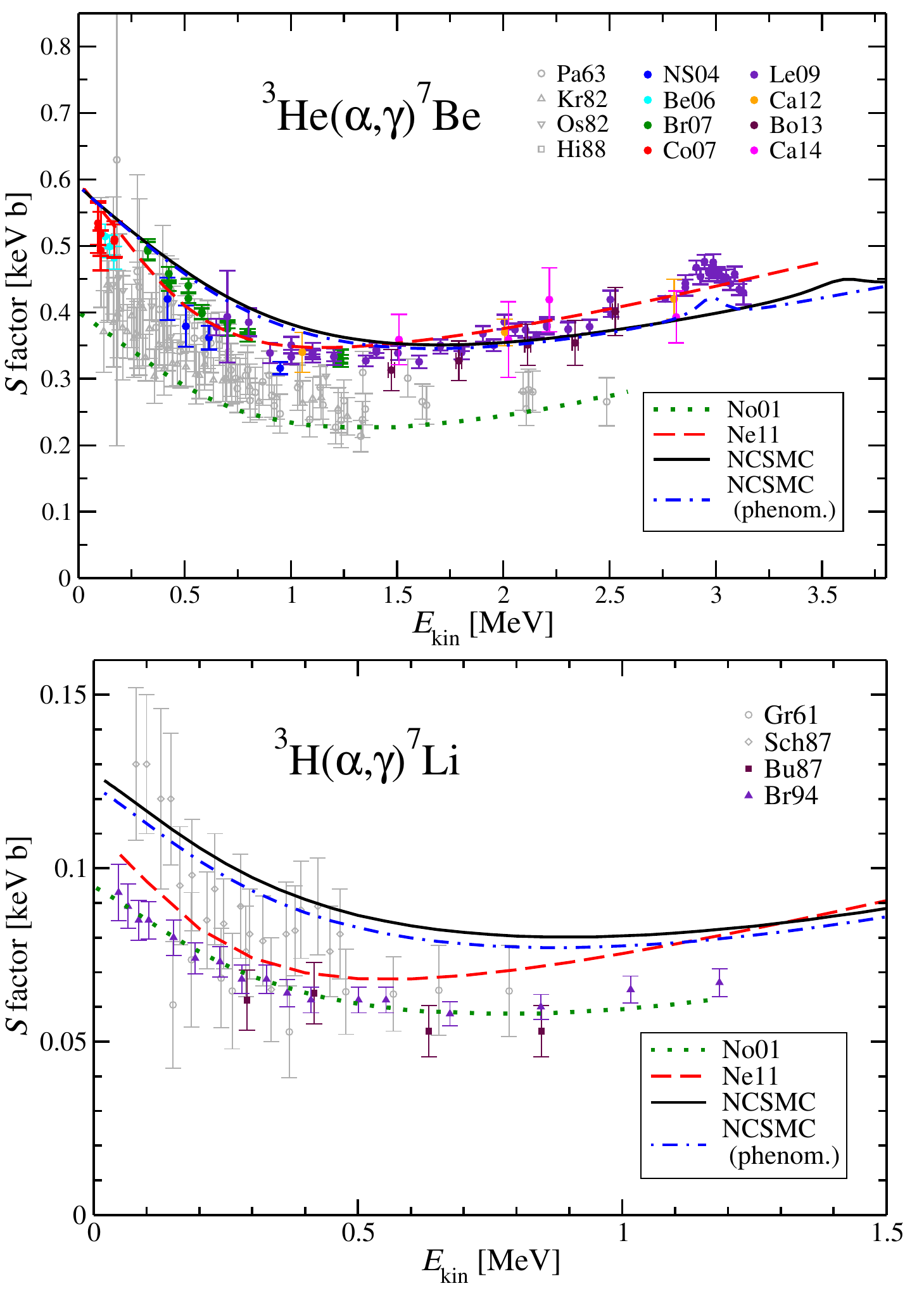}
     \end{subfigure}
            \hfill
    \begin{subfigure}[b]{0.49\textwidth}
         \centering
         \includegraphics[width=\textwidth]{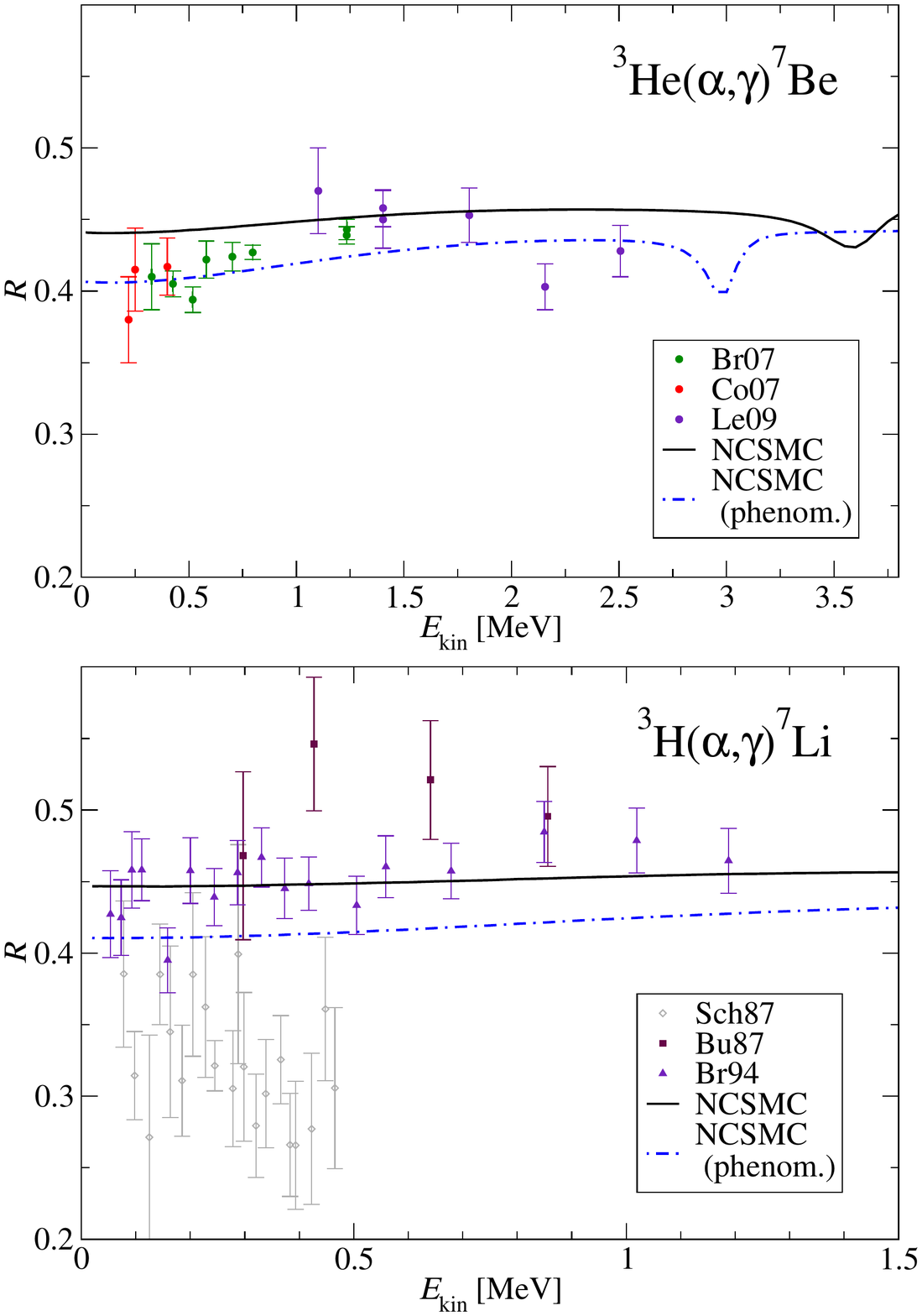}
     \end{subfigure}
    \caption{Left panels: Astrophysical S factor for the $^3$He($\alpha,\gamma$)$^7$Be and $^3$H($\alpha,\gamma$)$^7$Li radiative-capture processes obtained from the NCSMC approach and from its phenomenological version compared with other theoretical approaches and with experiments. Right panels: Ratio ($R$) of the $^3$He($\alpha,\gamma$)$^7$Be and $^3$H($\alpha,\gamma$)$^7$Li radiative-capture cross sections to the seven-nucleon excited state and to the seven-nucleon ground state obtained from the NCSMC, from its phenomenological version and from experiments. Recent data are in color and old data are in light grey. Adopted from Ref.~\cite{DohetEraly2016} where further details are given.}
    \label{fig:7Be7Li_Sfact}
\end{figure}

The ${^3{\rm He}}(\alpha,\gamma){^7{\rm Be}}$ and ${^3{\rm H}}(\alpha,\gamma){^7{\rm Li}}$ astrophysical S factors obtained with the NCSMC approach and with its phenomenological version are displayed in the left panels of Fig.~\ref{fig:7Be7Li_Sfact} and compared with experiment. The electric E1 and E2 transitions as well as the magnetic M1 transitions have been considered. 
For the energy ranges which are considered, the contribution of the E1 transitions is dominant while $M1$ contribution is essentially negligible and the E2 transitions play a small but visible role in the ${^3{\rm He}}(\alpha,\gamma){^7{\rm Be}}$ radiative capture, mostly near the $7/2^-$ resonance energy. Qualitatively, the ${^3{\rm He}}(\alpha,\gamma){^7{\rm Be}}$ astrophysical S factors agree rather well with the experimental ones. The results obtained with the phenomenological version are similar up to approximately the $7/2^-$ resonance energy. The peak in the experimental S factor at a relative collision energy of about 3~MeV corresponds to a E2 transition from the $7/2^-$ resonance to the $3/2^-$ ground state. Since the $7/2^-$ resonance energy is slightly overestimated by our theoretical approach, the energy of the corresponding peak in the S factor is also overestimated. This is corrected in the NCSMC-pheno calculation that utilizes the NCSM input energies as adjustable parameters. In the left panel of Fig.~\ref{fig:7Be7Li_Sfact}, other theoretical results based on two different realistic NN interactions~\cite{Nollett2001b,Neff2011} are also presented. Although the Fermion Molecular Dynamics approach of Ref.~\cite{Neff2011} is not fully able to describe the short-range correlations of the wave function, a good agreement between theoretical and experimental ${^3{\rm He}}(\alpha,\gamma){^7{\rm Be}}$ astrophysical S factor was obtained. 

The ${^3{\rm H}}(\alpha,\gamma){^7{\rm Li}}$ astrophysical S factor is overestimated over the full energy range in the NCSMC calculation. The phenomenological approach improves only slightly the situation. As shown in the left panel of Fig.~\ref{fig:7Be7Li_Sfact}, a similar behavior is present, though less pronounced, in calculations of Ref.~\cite{Neff2011} while the approach of Ref.~\cite{Nollett2001b} reproduces the ${^3{\rm H}}(\alpha,\gamma){^7{\rm Li}}$ astrophysical S factor but underestimates the ${^3{\rm He}}(\alpha,\gamma){^7{\rm Be}}$ one. 
This suggests a possible underestimation of the experimental systematic uncertainties and underscores the need for new experimental studies of ${^3{\rm H}}(\alpha,\gamma){^7{\rm Li}}$.

Finally, in the right panels of Fig.~\ref{fig:7Be7Li_Sfact}, we compare the ratio of the radiative-capture cross sections to the seven-nucleon excited state and to the seven-nucleon ground state obtained from the NCSMC approach and from experiments. Theoretical results agree rather well with the experimental data. Differences between the NCSMC approach and its phenomenological version are comparable with the size of the experimental error bars.

A further improvement of the ${^3{\rm He}}(\alpha,\gamma){^7{\rm Be}}$ and ${^3{\rm H}}(\alpha,\gamma){^7{\rm Li}}$ astrophysical S factors would require the inclusion of chiral as well as SRG-induced three-body forces in the NCSMC. Calculations in this direction are under way.

\subsection{$^7$Be(p,$\gamma$)$^8$B Radiative Capture}

The core temperature of the Sun can be determined with high accuracy through measurements of the $^8$B neutrino flux, currently known with a $\sim9\%$ precision~\cite{PhysRevLett.92.181301}. 
An important input in modeling this flux are the rates of the $^3$He($\alpha,\gamma$)$^7$Be and the $^7$Be(p,$\gamma$)$^8$B radiative capture reactions~\cite{RevModPhys.83.195,RevModPhys.70.1265}. The $^7$Be(p,$\gamma$)$^8$B reaction constitutes the final step of the nucleosynthetic chain leading to $^8$B. At solar energies this reaction proceeds by external, predominantly nonresonant E1, $S$- and $D$-wave capture into the weakly-bound $2^+$ ground state of $^8$B (bound only by 137.5 keV with respect to the $^7$Be+p threshold). Experimental determinations of the $^7$Be(p,$\gamma$)$^8$B capture include direct measurements with proton beams on $^7$Be targets~\cite{PhysRevLett.50.412,PhysRevC.68.065803} as well as indirect measurements through the breakup of a $^8$B projectile into $^7$Be and proton in the Coulomb field of a heavy target~\cite{PhysRevC.73.015806}. Theoretical calculations needed to extrapolate the measured S factor to the astrophysically relevant Gamow-window energy were performed with several methods: the R-matrix parametrization~\cite{Barker1995693}, the potential model~\cite{PhysRevC.68.045802}, microscopic cluster models~\cite{PhysRevC.70.065802} and also using the {\it ab initio} no-core shell model wave functions for the $^8$B bound state~\cite{Navratil2006}, as well as by an application of Bayesian methods in combination with the Halo Effective Field Theory~\cite{Zhang2015}. The most recent evaluation of the $^7$Be(p,$\gamma$)$^8$B astrophysical S factor at zero energy, $S_{17}(0)$, has a $\sim$10\% error dominated by the uncertainty in theory~\cite{RevModPhys.83.195}. 

\begin{figure}
    \centering
    \begin{subfigure}[b]{0.49\textwidth}
         \centering
         \includegraphics[width=\textwidth]{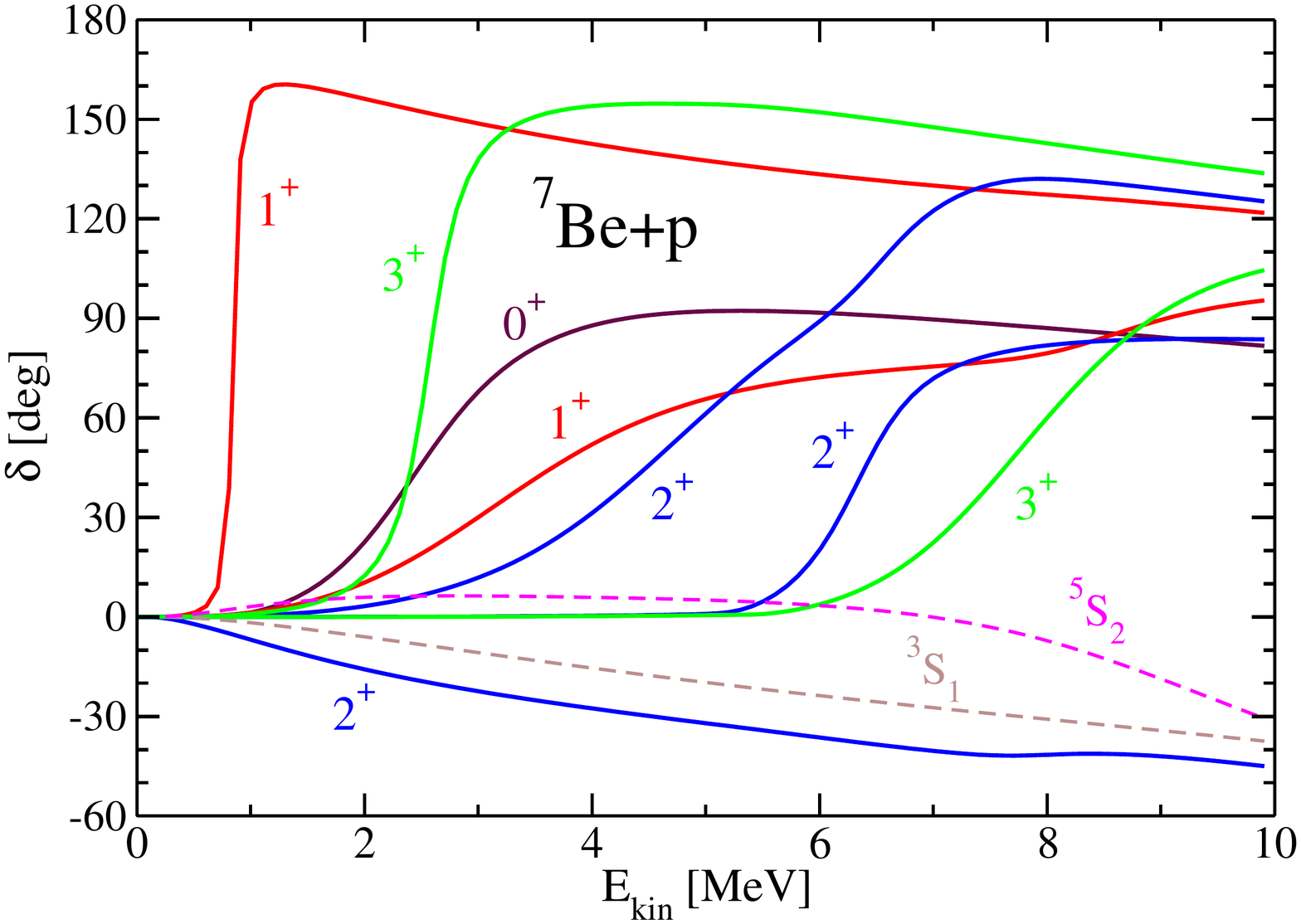}
    %\caption{Adopted from Ref.~\cite{Kravvaris2022ab} where further details are given.}
    %\label{fig:8B_phase}
     \end{subfigure}
            \hfill
    \begin{subfigure}[b]{0.49\textwidth}
         \centering
         \includegraphics[width=\textwidth]{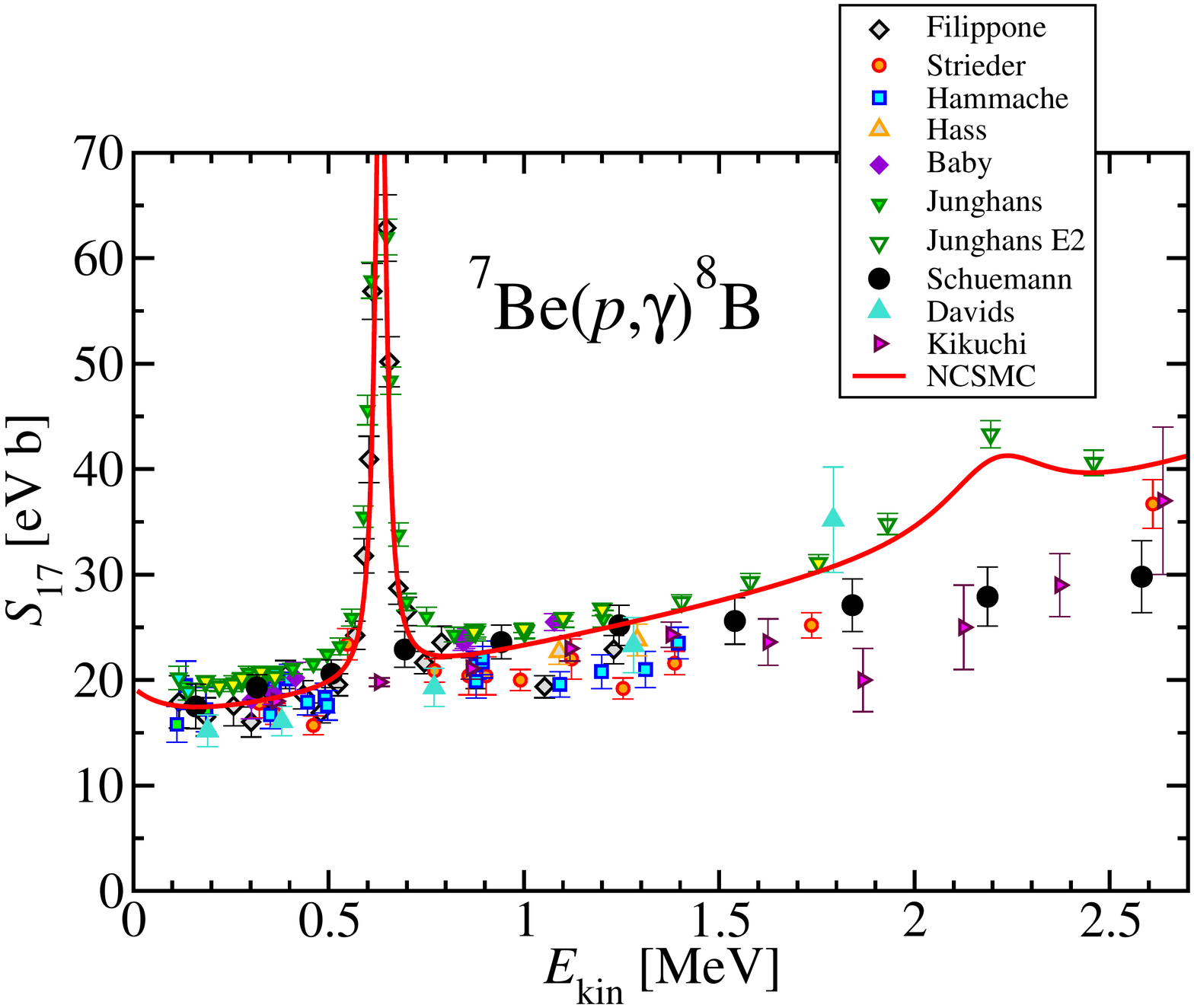}
    %\caption{Adopted from Ref.~\cite{Kravvaris2022ab} where further details are given.}
    %\label{fig:8B_Sfact}
     \end{subfigure}
     \caption{Left panel: $^7$Be+p eigenphase shifts (solid lines) and $^3S_1$ and $^5S_2$ diagonal phase shifts (dashed lines) obtained from the NCSMC approach. Right panel: Astrophysical S factor of the $^7$Be(p,$\gamma$)$^8$B radiative capture obtained from the NCSMC-pheno approach compared to experimental data. $E_{\rm kin}$ is the energy in the c.m. frame. Adopted from Ref.~\cite{Kravvaris2022ab} where further details are given.}
    \label{fig:8B}
\end{figure}

The first {\it ab initio} many-body calculations of the $^7$Be(p,$\gamma$)$^8$B capture starting from a nucleon-nucleon interaction that describes two-nucleon properties with high accuracy were reported in Ref.~\cite{Navratil2011a}. Those calculations were performed within the NCSM/RGM formalism using an SRG evolved chiral N$^3$LO NN~\cite{Entem2003} with the SRG evolution parameter $\Lambda{=}1.86\; {\rm fm}^{-1}$ chosen so that the experimental separation energy of the $^8$B weakly bound $2^+$ g.s. with respect to the $^7$Be+p is reproduced in the largest-space calculation that was reached. More advanced calculations have now been performed within the NCSMC formalism using a set of chiral NN and 3N interactions~\cite{Kravvaris2022ab}. Unlike the NCSM/RGM calculations focusing only on the direct E1 capture, the new NCSMC calculations also include the M1 and E2 contributions from resonances. As an example of obtained results, we present in Fig.~\ref{fig:8B} the $^7$Be+p phase shifts and the astrophysical S factor obtained using the chiral NN interaction from Ref.~\cite{Entem2017} and the chiral 3N interaction of the type introduced in Ref.~\cite{Soma2020} that in addition includes a sub-leading contact interaction enhancing the strength of the spin-orbit interaction~\cite{Girlanda2011}. The positive parity eigenphase shifts (obtained by the diagonalization of the multi-channel S-matrix) show the well-established $1^+_1$ and $3^+_1$ resonances as well as predictions of several other broader resonances. Unlike the NCSM/RGM calculations~\cite{Navratil2011a}, the NCSMC $S$-wave phase shifts shown in the left panel of Fig.~\ref{fig:8B} manifest consistent scattering length signs with those determined in recent measurements (negative for $^5S_2$, positive for $^3S_1$)~\cite{PhysRevC.99.045807}. The calculated astrophysical S factor shown in the right panel of Fig.~\ref{fig:8B} was obtained within the NCSMC-pheno approach. It reproduces well the resonance contributions due to the M1 and to a smaller extent to the E2 transitions from the $1^+$ resonance (the sharp peak at $\sim0.6$~MeV) and $3^+$ resonance (the bump at $\sim2.2$~MeV) to the weakly bound $2^+$ ground state of $^8$B; compare the peak positions to the eigenphase shifts in the left panel that are slightly shifted as they were obtained within the original NCSMC, i.e., without any phenomenological adjustments. The calculated S factor matches well the Junghans direct measurement data~\cite{PhysRevC.68.065803} starting at the $1^+$ resonance in the whole displayed range including the $3^+$ bump. At low energies below the $1^+$ resonance, the NCSMC-pheno results are slightly below the Junghans data. Overall, the new NCSMC calculations ~\cite{Kravvaris2022ab} are consistent with the latest recommended S factor value at zero energy: $S_{17}(0)=20.8\pm0.7\mathrm{(expt)}\pm1.4\mathrm{(theory)}$ eV$\cdot$barn~\cite{RevModPhys.83.195}. However, the theoretical uncertainty is reduced by more than a factor of 5. 

\subsection{$^8$Li(n,$\gamma$)$^9$Li Radiative Capture}

In neutron rich astrophysical environments, reactions involving the short-lived $^8$Li nucleus 
may contribute to the synthesis of heavier nuclei by bridging the stability gap of mass $A{=}8$ elements. In particular, the $^8$Li(n,$\gamma$)$^9$Li capture reaction plays an important role in inhomogeneous Big Bang Nucleosynthesis and in the r-process. 
There, it competes with the $^8$Li($\alpha$,n)$^{11}$B reaction and the $^8$Li beta decay, affecting the reaction path to $A{>}8$ isotopes and also the abundances of Li, Be, B, and C. The relevant reaction chains are $^7$Li(n,$\gamma$)$^8$Li($\alpha$,n)$^{11}$B(n,$\gamma$)$^{12}$B($\beta^+$)$^{12}$C and $^7$Li(n,$\gamma$)$^8$Li(n,$\gamma$)$^9$Li($\alpha$,n)$^{12}$B($\beta^+$)$^{12}$C~\cite{PhysRevLett.68.1283,Rauscher1994}. In addition, the reaction chain with two-neutron captures \newline $^4$He(2n,$\gamma$)$^6$He(2n,$\gamma$)$^8$He($\beta^-$)$^8$Li(n,$\gamma$)$^9$Li($\beta^-$)$^9$Be, of which the $^8$Li(n, $\gamma$)$^9$Li is also a component, has been considered as an alternative to the triple-alpha process in overcoming the $A=8$ mass gap in the r-process 
for supernovae of type II~\cite{PhysRevC.52.2231,Efros1996}. 

\begin{figure}
    \centering
    \begin{subfigure}[b]{0.49\textwidth}
         \centering
         \includegraphics[width=\textwidth]{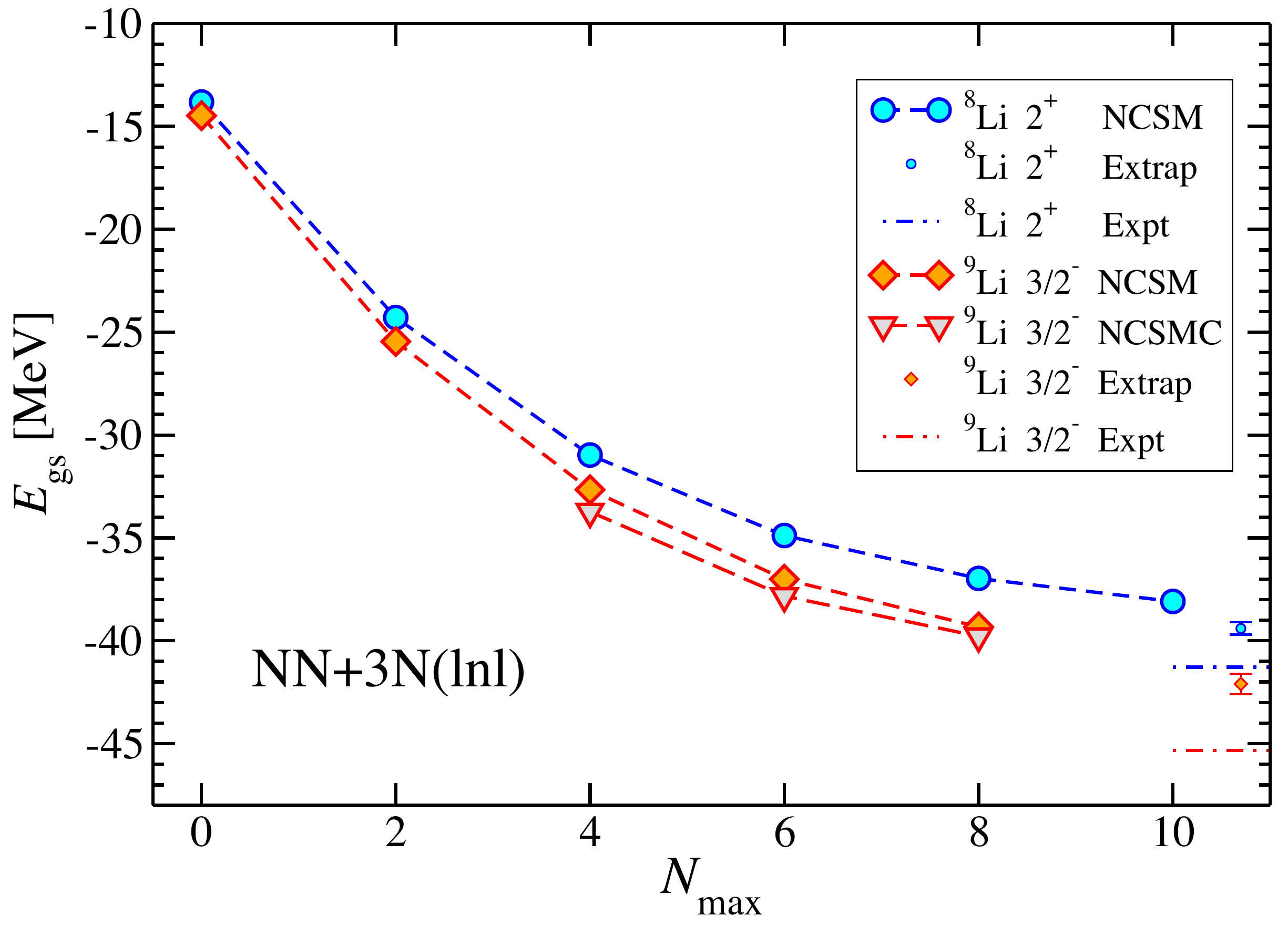}
    %\caption{Adopted from Ref.~\cite{PhysRevC.103.035801} where further details are given.}
    %\label{fig:8Li9Li_gs}
     \end{subfigure}
            \hfill
    \begin{subfigure}[b]{0.49\textwidth}
         \centering
         \includegraphics[width=\textwidth]{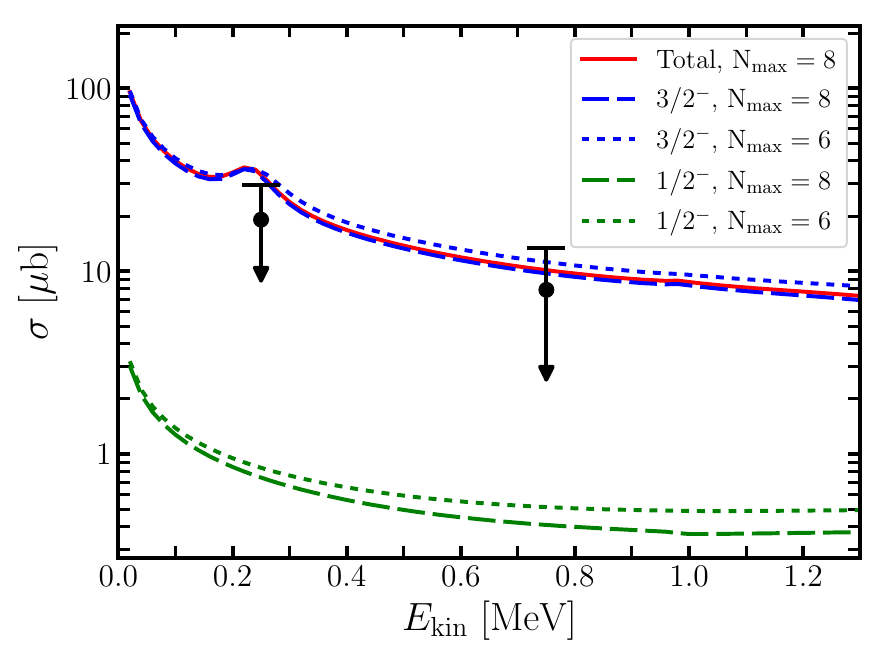}
    %\caption{Adopted from Ref.~\cite{PhysRevC.103.035801} where further details are given.}
    %\label{fig:9Li_xsect}
     \end{subfigure}
     \caption{Left panel: $^8$Li (circles) and $^9$Li (diamonds) ground-state energy
dependence on the size of the NCSM and for $^9$Li also NCSMC (triangles) basis. Extrapolations to infinite $N_{\rm max}$ with their uncertainties are presented on the right. The experimental values are shown by dashed-dotted lines. Right panel: $^8$Li(n,$\gamma$)$^9$Li radiative capture cross section obtained in the NCSMC-pheno calculations. We compare $N_{\rm max}{=}6$ (dotted lines), $N_{\rm max}{=}8$ (dashed lines), the total $N_{\rm max}{=}8$ cross-section (solid line), and experimental limits from Ref.~\cite{PhysRevC.57.959} (black points). Cross-section contributions from the ground state are shown in blue, contributions from the first excited state are in green. Adopted from Ref.~\cite{PhysRevC.103.035801} where further details are given.}
    \label{fig:9Li_gs_xsect}
\end{figure}

As the $^8$Li half-life is 840 ms and a neutron target is not available, the $^8$Li(n,$\gamma$)$^9$Li reaction cannot be measured directly. There have been several attempts to determine its cross section by indirect methods, e.g., as using a radioactive beam of $^9$Li and the Coulomb-dissociation method with U and Pb targets ~\cite{PhysRevC.57.959} or by measuring transfer reactions with the radioactive $^8$Li beam. The $^8$Li($n,\gamma$)$^9$Li cross section and its reaction rate have also been the focus of several theoretical investigations, based on various approaches including the shell model, potential model, and the microscopic cluster model. 

The first {\it ab initio} investigation of this reaction was reported in Ref.~\cite{PhysRevC.103.035801}. The NCSMC approach was applied with the input consisting of the chiral NN interaction~\cite{Entem2003} and the chiral 3N from Ref.~\cite{Soma2020}. In the left panel of Fig.~\ref{fig:9Li_gs_xsect}, we present the g.s. energy dependence on the basis size characterized by $N_{\rm max}$ obtained within the NCSM for both $^8$Li and $^9$Li. For the latter, we also show the NCSMC results for $N_{\rm max}{=}4, 6$, and 8. It should be noted that the NCSM calculations serve as input for the NCSMC. NCSMC calculations increase the binding energies compared to the NCSM results at any fixed $N_{\rm max}$ due to the inclusion of the cluster basis component. In the same figure, we show in addition the NCSM g.s. energies extrapolated to infinite basis size using the exponential function $E (N_{\rm max}) = E_{\infty} + a e^{- b N_{\rm max}}$ with an uncertainty obtained by varying the number of $N_{\rm max}$ points (see also the discussion of Fig.~\ref{fig:6He_gs}). The extrapolated energies overestimate experiment by a few percent.

\begin{figure}
    \centering
    \includegraphics[width=\textwidth]{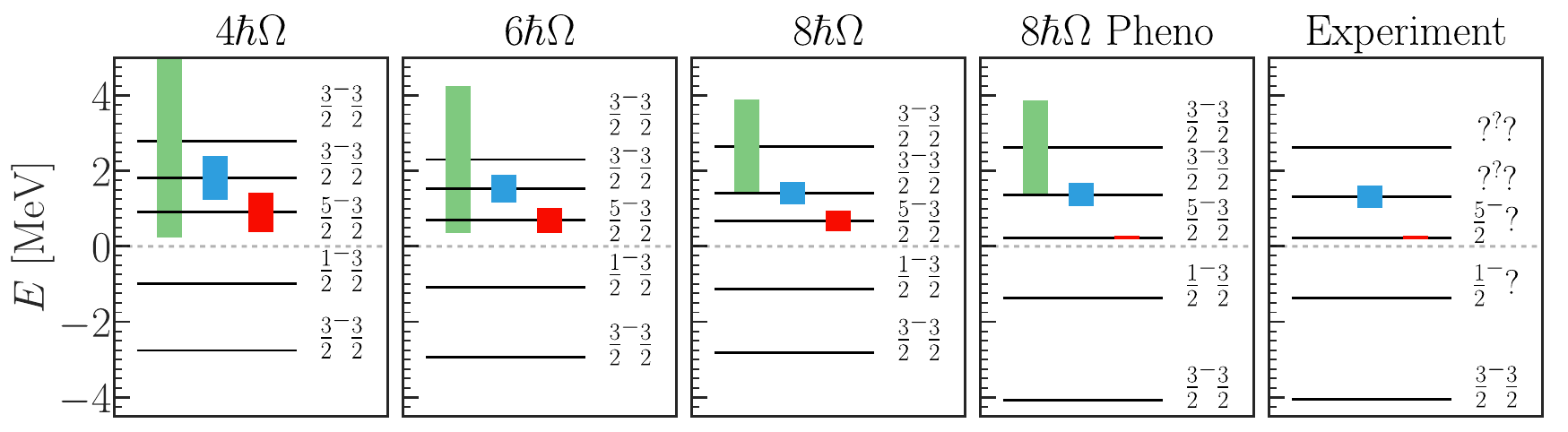}
    \caption{Energies of $^9$Li bound states and low-lying resonances with respect to the $^8$Li+n threshold. The leftmost three panels show NCSMC calculations at $N_{\rm max}{=}4, 6$, and 8. The fourth panel shows the NCSMC-pheno $N_{\rm max}{=}8$ calculation. Coloured bars represent the widths of resonances. The rightmost panel shows the experimental data. Question marks are used where data is unavailable. Adopted from Ref.~\cite{PhysRevC.103.035801} where further details are given.}
    \label{fig:9Li_levels}
\end{figure}

In the leftmost three panels of Fig.~\ref{fig:9Li_levels}, we show the NCSMC results for $^9$Li bound-state energies and energies and widths of three resonances obtained in the $N_{\rm max}{=}4,6,8$ model spaces. These are shown alongside available experimental data. It is clear that the calculated $5/2^-$ resonance is a good match to the experimentally known resonance at 4.296 MeV. We predict that the experimental 5.38 MeV level is $3/2^-$. On the other hand, the experimentally very narrow 6.43 MeV level does not correspond to our calculated very broad second $3/2^-$. Rather, it presumably corresponds to the calculated $7/2^-$ state (not shown in the figure). The NCSMC-pheno results in the fourth panel predict the width of the $5/2^-$ and the $3/2^-$ resonances in a very good agreement with experiment (assuming that the 5.38 MeV level is $3/2^-$).

The calculated $^8$Li(n,$\gamma$)$^9$Li capture cross section is presented in the right panel of Fig.~\ref{fig:9Li_gs_xsect}. NCSMC-pheno results obtained in the $N_{\rm max}{=}8$ and $N_{\rm max}{=}6$ spaces are compared. Overall, we find a good stability of the calculations. By increasing the model space, the cross section gets reduced slightly and the difference can serve as an estimate of the uncertainty. The capture to the $^9$Li ground state dominates the total cross section. The excited state contribution is suppressed by more than an order of magnitude. In the low-energy region displayed in the figure, the non-resonant E1 capture is the leading contribution. The E2/M1 capture enhanced by the $5/2^-$ resonance is visible as a bump around 0.23 MeV.

The NCSMC calculated cross section is on the higher side but still within the limits derived from the 1998 NSCL Coulomb dissociation experiment~\cite{PhysRevC.57.959} shown in the left panel of Fig.~\ref{fig:9Li_gs_xsect} by black points and vertical lines. These limits should be compared to the E1 contribution to the capture to the ground state. The reaction rate obtained from the NCSMC calculated capture cross section is lower than early evaluations. However, it is higher by about a factor of two compared to recent potential cluster model calculations. These results indicate that the $^8$Li(n,$\gamma$)$^9$Li reaction might play a more important astrophysical role than recently considered.  

\section{Concluding Remarks}

{\it Ab initio} theory of light and medium mass nuclei is a rapidly evolving field with many exciting advances in the past decade. Several new methods have been introduced capable to describe bound-state properties of nuclei as heavy as tin and beyond. Similarly, there has been a significant progress in calculations of unbound states, nuclear scattering and reactions, mostly in light nuclei so far.

In this chapter, we discussed the recently introduced unified approach to nuclear bound and continuum states based on the coupling of a square-integrable basis ($A$-nucleon NCSM eigenstates), suitable for the description of many-nucleon correlations, and a continuous basis (NCSM/RGM cluster states) suitable for a description of long-range correlations, cluster correlations and scattering. 
This {\it ab initio} method, the no-core shell model with continuum, is capable of describing efficiently: $i)$ short- and medium-range nucleon-nucleon correlations thanks to the large HO basis expansions used to obtain the NCSM eigenstates, and $ii)$ long-range cluster correlations thanks to the NCSM/RGM cluster-basis expansion. 

Results presented in this chapter demonstrate current capabilities of the NCSMC. With high-precision chiral NN+3N interactions as the input, we are able to predict with confidence properties of light nuclei even with a large neutron or proton excess. We focused in particular on applications to reactions important for astrophysics and reviewed results for the DT and D$^3$He fusion, $^3$He($\alpha,\gamma$)$^7$Be, $^3$H($\alpha,\gamma$)$^7$Li, $^7$Be(p,$\gamma$)$^8$B, and $^8$Li(n,$\gamma$)$^9$Li radiative captures. NCSMC calculations of several other radiative capture reactions important for astrophysics including $^2$H($\alpha,\gamma$)$^6$Li, $^{11}$C(p,$\gamma$)$^{12}$N, and $^{14}$C(n,$\gamma$)$^{15}$C have just been completed~\cite{Hebborn:2022iiz} or are under way. In the future we plan on quantifying the uncertainty related to the convergence of the chiral expansion (before and after the application of phenomenological corrections) by extending the methodology of Ref.~\cite{PhysRevC.102.024616} to capture cross sections.

Long-term goal is to study $\alpha$ clustering within the NCSMC, e.g., in $^{12}$C and $^{16}$O, and reactions involving $^4$He, e.g., $^8$Be($\alpha,\gamma$)$^{12}$C, $^{12}$C($\alpha,\gamma$)$^{16}$O important for stellar burning, the $^{11}$B($p,\alpha$)$^8$Be aneutron reaction explored as a candidate for the future fusion energy generation as well as the $^{13}$C($\alpha,n$)$^{16}$O relevant to the i- and s-processes. First steps in this direction have already been taken~\cite{Kravvaris:2020cvn}.

\section{Acknowledgments}

We thank all our collaborators who contributed to the development of NCSMC method, in particular Guillaume Hupin, Kostas Kravvaris, Simone Baroni, Carolina Romero-Redondo, J\'er\'emy Dohet-Eraly, and Angelo Calci. This work was supported by the NSERC Grants No. SAPIN-2016-00033 and SAPPJ-2019-00039 and by the U.S. Department of Energy, Office of Science, Office of Nuclear Physics, under Work Proposal No. SCW0498. TRIUMF receives federal funding via a contribution agreement with the National Research Council of Canada. This work was prepared in part by LLNL under Contract No. DE-AC52-07NA27344. Computing support came from an INCITE Award on the Summit supercomputer of the Oak Ridge Leadership Computing Facility (OLCF) at ORNL, from the LLNL institutional Computing Grand Challenge program, from Westgrid and Compute Canada.

%\input{referenc}
%\bibliography{mybib}
%\bibliographystyle{unsrt}
%\bibliographystyle{aasjournal}

\end{document}